\documentclass{aastex631}

\shorttitle{Predicting SEPs Using \textit{SDO}/HMI Vector Magnetic Data Products}
\shortauthors{Abduallah et al.}

\usepackage{pgfplotstable}
\usepackage{pgfplots}
\pgfplotsset{compat=1.8}
\usetikzlibrary{matrix,calc,matrix.skeleton}
\DeclareMathAlphabet {\mathbfit}{OML}{cmm}{b}{it}

\begin{document}

\title{{Predicting Solar Energetic Particles Using \textit{SDO}/HMI Vector Magnetic Data Products and a Bidirectional LSTM Network}}

\author{Yasser Abduallah}
\affiliation{Institute for Space Weather Sciences,
New Jersey Institute of Technology,
University Heights, Newark, NJ 07102, USA;
ya54@njit.edu, wangj@njit.edu, haimin.wang@njit.edu}
\affiliation{Department of Computer Science,
New Jersey Institute of Technology, University Heights, Newark, NJ 07102, USA}

\author{Vania K. Jordanova}
\affiliation{Space Science and Applications, Los Alamos National Laboratory, Los Alamos, NM 87545, USA}

\author{Hao Liu}
\affiliation{Institute for Space Weather Sciences,
New Jersey Institute of Technology,
University Heights, Newark, NJ 07102, USA;
ya54@njit.edu, wangj@njit.edu, haimin.wang@njit.edu}

\author{Qin Li}
\affiliation{Institute for Space Weather Sciences, 
New Jersey Institute of Technology, 
University Heights, Newark, NJ 07102, USA;
ya54@njit.edu, wangj@njit.edu, haimin.wang@njit.edu}
\affiliation{Big Bear Solar Observatory, New Jersey Institute of Technology, 40386 North Shore Lane, Big Bear City, CA 92314, USA}
\affiliation{Center for Solar-Terrestrial Research, New Jersey Institute of Technology, University Heights, Newark, NJ 07102, USA}

\author{Jason T. L. Wang}
\affiliation{Institute for Space Weather Sciences,
New Jersey Institute of Technology,
University Heights, Newark, NJ 07102, USA;
ya54@njit.edu, wangj@njit.edu, haimin.wang@njit.edu}
\affiliation{Department of Computer Science,
New Jersey Institute of Technology, University Heights, Newark, NJ 07102, USA}

\author{Haimin Wang}
\affiliation{Institute for Space Weather Sciences,
New Jersey Institute of Technology,
University Heights, Newark, NJ 07102, USA;
ya54@njit.edu, wangj@njit.edu, haimin.wang@njit.edu}
\affiliation{Big Bear Solar Observatory, New Jersey Institute of Technology, 40386 North Shore Lane, Big Bear City, CA 92314, USA}
\affiliation{Center for Solar-Terrestrial Research, New Jersey Institute of Technology, University Heights, Newark, NJ 07102, USA}

\begin{abstract}

Solar energetic particles (SEPs) are an essential source of space radiation, 
which are hazards for humans in space, spacecraft, and technology in general. 
In this paper we propose a deep learning method, specifically a bidirectional long short-term memory (biLSTM) network,
to predict if an active region (AR) would produce an SEP event given that 
(i) the AR will produce 
an M- or X-class flare and a coronal mass ejection (CME)
associated with the flare, or
(ii) the AR will produce 
an M- or X-class flare
regardless of whether or not the flare is associated with a CME.
The data samples used in this study are collected from the \textit{Geostationary Operational Environmental Satellite}'s X-ray flare catalogs 
provided by the National Centers for Environmental Information.
We select M- and X-class flares
with identified ARs in the catalogs for the period between 
2010 and 2021, and
find the associations of flares, CMEs and SEPs in the Space Weather Database of Notifications, Knowledge, Information
during the same period. 
Each data sample contains physical parameters collected from the Helioseismic and Magnetic Imager on board
the \textit{Solar Dynamics Observatory}. 
Experimental results based on different performance metrics
demonstrate that the proposed biLSTM network is better than related machine learning algorithms for the two SEP prediction tasks studied here.
We also discuss extensions of our approach for probabilistic forecasting and calibration with empirical evaluation.

\end{abstract}
\keywords{Solar energetic particles; Coronal mass ejections; Solar flares; Solar activity}

\section{Introduction} \label{sec:intro}

Solar eruptions including flares and coronal mass ejections (CMEs) can endanger modern civilization. 
Solar flares are large bursts of radiation released into space;
they appear as sudden and unexpected brightening in the solar atmosphere with a duration ranging from minutes to hours. 
CMEs are significant discharges of plasma and magnetic fields 
produced by the solar corona into the interplanetary medium \citep{CMEEffect2000JGR...105.2375L}. 
They are considered to be the largest-scale solar eruptions in the solar system and occur on a quasi-regular basis 
\citep{CMEEffectChecn2011LRSP....8....1C,CMEEffectWebbHoward2012LRSP....9....3W,CMEEffectKiplua2017LRSP...14....5K}. 
Research shows that both flares and CMEs are magnetic events, sharing a similar physical process 
\citep{FlaresCMESSimilar1995AA...304..585H,FlareCMESimRelation2012ApJ...753...88B}, 
though more work is performed to understand the correlation between them 
\citep{FlaresCMEInvestigationYashiro_gopalswamy_2008,FlaresCMEInvestigation2018ApJ...869...99K}. 
Large flares and accompanied CMEs cause solar energetic particles (SEPs).
SEPs, composed of electrons, protons and heavy ions, 
are expedited by magnetic reconnection or shock waves associated with the CMEs
\citep{2018JASTP.177..131B,SEPRelFlareFlux2018RNAAS}.
When SEP events are strong, they cause nuclear cascades in the Earth's upper atmosphere
and also represent a radiation hazard to equipment in space that is not adequately protected 
\citep{SEPDistributionReames2013SoPh..285..233R,2018JASTP.177..148J,RJ-2020}.

Active regions (ARs), which manifest complex magnetic geometry and properties \citep{ARExibitBenz2008LRSP....5....1B},
are the source of flares and CMEs 
\citep{CMEEffectChecn2011LRSP....8....1C,AREvolutionVanDriel2015LRSP...12....1V}. 
The lifetime of ARs ranges from days to months \citep{AREvolutionVanDriel2015LRSP...12....1V}.
Recently, researchers combine machine learning with physical parameters derived from vector magnetograms 
provided by the Helioseismic and Magnetic Imager
\citep[HMI;][]{HMIPolarizationcaliSchou2012SoPh..275..327S} on board the 
\textit{Solar Dynamics Observatory} 
\citep[\textit {SDO};][]{SDOPesnell2012}
to predict flares, CMEs, and SEPs. 
These physical parameters, including magnetic helicity and magnetic flux 
\citep{FlaresMagenticFluxLeka2003ApJ...595.1277L,FlaresMagenticFluxSchrijver2007ApJ...655L.117S,FlaresMagenticFluxMoore2012ApJ...750...24M},
are part of the vector magnetic data products, named the Space-weather HMI Active Region Patches \citep[SHARP;][]{Bobra:2014SoPh..289.3549B}, 
produced by the \textit{SDO}/HMI team. 

Machine learning (ML) has been popular in predictive analytics for many years. 
ML is able to learn patterns from historical data and make predictions on unseen or future data~
\citep{MLNewAIAlpaydin2016,Goodfellow_DeepLearningBookDBLP:books/daglib/0040158}.
For example, \citet{Liu..Wang..Solar..2017ApJ...843..104L} used random forests (RF) and the SHARP parameters to 
predict the occurrence of a certain class of flares in a given AR within 24 hours. 
\citet{2018SoPh..293...48J} employed machine learning to extract relevant information 
from photospheric and coronal image data to perform flare prediction. 
\citet{ForecastingFlaresPredictors2018SoPh..293...28F} adopted multiple machine learning algorithms including RF, 
multilayer perceptrons (MLP) and support vector machines (SVM) for flare forecasting.
More recently, researchers started to use deep learning (DL),
which is a branch of machine learning focusing on the use of deep neural networks, to enhance the learning outcome
\citep{Goodfellow_DeepLearningBookDBLP:books/daglib/0040158}.
\citet{DeepLearningFlareForecastingLoS2018ApJ...856....7H} designed a convolutional neural network 
to learn patterns from line-of-sight magnetograms of ARs and
used the patterns to forecast flares. \citet{Liu_2019FlarePrediction} adopted a long short-term memory (LSTM) network for flare prediction. 
\citet{CMH-2019} employed LSTM and the SHARP parameters to identify solar flare precursors;
the authors later extended their work by investigating solar cycle dependence \citep{WCT-2020}.
Similar ML and DL methods have been applied to CME and SEP forecasting.
\citet{BobraCME2016ApJ...821..127B} used SVM to predict CMEs;
\citet{Liu_2020CMEPrediction} extended their work by adopting recurrent neural networks including LSTM and gated recurrent units.
\citet{Inceoglu_2018CMESEP} employed SVM and MLP 
to forecast if flares would be accompanied with CMEs and SEPs. 

In this paper, we propose a new deep learning method,
specifically a bidirectional long short-term memory (biLSTM) network,
for SEP prediction using the \textit{SDO}/HMI vector magnetic data products. 
We aim to solve two binary prediction problems:
(i) predict whether an AR would produce an SEP event given that 
the AR will produce 
an M- or X-class flare and a CME associated with the flare
(referred to as the FC\_S problem);
(ii) predict whether an AR would produce an SEP event given that 
the AR will produce 
an M- or X-class flare
regardless of whether or not the flare is associated with a CME 
(referred to as the F\_S problem).
The proposed biLSTM is an extension of LSTM \citep{LSTMHochreiter1997LongSM}, 
both of which are well suited for time series forecasting
\citep{DeepLearningBook2LeCun2015,Goodfellow_DeepLearningBookDBLP:books/daglib/0040158}.
Unlike LSTM, which works in one direction,
biLSTM works back and forth on the input data
and then the patterns learned from the two directions are joined together to strengthen the learning outcome.
In SEP prediction, the observations and physical parameters associated with ARs
form time series, and hence biLSTM is suitable for our study.

The rest of this paper is organized as follows. 
Section \ref{sec:data} explains the data and data collection procedure used in our study.
Section \ref{sec:methodology} describes our proposed deep learning method.
Section \ref{sec:result} reports experimental results and
discusses extensions of our approach for probabilistic forecasting and calibration.
Section \ref{sec:conclusions} concludes the paper.

\vspace{6pt}
\section{Data} 
\label{sec:data}
In this work we adopt the Space-weather HMI Active Region Patches \citep[SHARP;][]{Bobra:2014SoPh..289.3549B}
that were produced by the \textit{SDO}/HMI team and released at the end of 2012.
These data are available for download, 
in the data series {\sf hmi.sharp},
from the Joint Science Operations Center (JSOC).\footnote{\url{http://jsoc.stanford.edu/}} 
The SHARP data provide physical parameters of active regions (ARs)
that have been used to predict flares, CMEs and SEPs
\citep{BobraCME2016ApJ...821..127B, Liu..Wang..Solar..2017ApJ...843..104L,Inceoglu_2018CMESEP,Liu_2019FlarePrediction,Liu_2020CMEPrediction}.
We collected SHARP data samples from the data series, {\sf hmi.sharp\_cea\_720s},
using the Python package SunPy \citep{SunPy2015CSD....8a4009S} at a cadence of 12 minutes. 
In collecting the data samples, we focused on the 18 physical parameters previously used for SEP prediction \citep{Inceoglu_2018CMESEP}.
These 18 SHARP parameters include the 
absolute value of the net current helicity (ABSNJZH),
area of strong field pixels in the active region (AREA\_AC),
mean characteristic twist parameter (MEANALP), 
mean angle of field from radial (MEANGAM), 
mean gradient of horizontal field (MEANGBH), 
mean gradient of total field (MEANGBT), 
mean gradient of vertical field (MEANGBZ), 
mean vertical current density (MEANJZD), 
mean current helicity (MEANJZH), 
mean photospheric magnetic free energy (MEANPOT), 
mean shear angle (MEANSHR), 
sum of flux near polarity inversion line (R\_VALUE), 
sum of the modulus of the net current per polarity (SAVNCPP), 
fraction of area with shear $>$ 45$^\circ$ (SHRGT45), 
total photospheric magnetic free energy density (TOTPOT), 
total unsigned current helicity (TOTUSJH), 
total unsigned vertical current (TOTUSJZ), 
and total unsigned flux (USFLUX). 

Since the 18 SHARP parameters have different units and scales, we normalized the parameter values 
using the min-max normalization procedure 
as done in \cite{Liu_2020CMEPrediction}. 
Each data sample contains the 18 SHARP parameters. 
Let $p^k_{i}$ be the original value of the $i$th parameter of the $k$th data sample.
Let $q^k_{i}$ be the normalized value of the $i$th parameter of the $k$th data sample.
Let $min_i$ be the minimum value of the $i$th parameter.
Let $max_i$ be the maximum value of the $i$th parameter.
Then
\begin{equation}
q^k_{i} = \frac{p^k_i - min_i}{max_i - min_i}.
\end{equation}

Appropriately labeling the data samples is crucial for machine learning.
We surveyed M- and X-class flares that occurred between 2010 and 2021 
with identified active regions in the \textit{GOES} X-ray flare catalogs provided by 
the National Centers for Environmental Information (NCEI). 
As done in \citet{BobraCME2016ApJ...821..127B}, we excluded 
ARs that were outside $\pm$ 70$^\circ$ of the central meridian 
because the SHARP parameters cannot be calculated correctly 
based on the vector magnetograms of the ARs that are near the limb
due to foreshortening and projection effects.\footnote{Notice that flaring ARs
outside $\pm$ 70$^\circ$ of the central meridian
may produce eruptive events that have increased probabilities to result in 
SEPs due to the magnetic connectivity with Earth. 
Excluding these flaring ARs may reduce the number of SEP events 
considered in the study.
This is a limitation of our approach.}
We also excluded
flares with an absolute value of the radial velocity of {\em SDO} being greater than 3500 m $s^{-1}$, 
low-quality HMI data as described by \citet{Hoeksema2014SoPh..289.3483HHMIQuality},
and data samples with incomplete SHARP parameters.
In this way we excluded data samples of low quality, and
kept qualified data samples of high quality in our study.
 Furthermore, we collected and extracted information from NASA's
Space Weather Database of Notifications, Knowledge, Information 
(DONKI)\footnote{\url{http://kauai.ccmc.gsfc.nasa.gov/DONKI/}}
to tag, for any given
M- or X-class flare, 
whether it produced a CME and/or SEP event.
We cross-checked the flare records in DONKI and \textit{GOES} X-ray flare catalogs
to ensure that each flare record was associated with an active region;
otherwise the flare record was removed from our study.

We then created two databases of active regions (ARs) for the period between 2010 and 2021.
ARs from 2010, 2016, and 2018-2021 were excluded from the study 
due to the lack of qualified data samples or
the absence of SEP events associated with M-/X-class flares and CMEs.
Thus, the databases contain ARs from six years, namely 2011-2015 and 2017.
In our first database, referred to as the FC\_S database,
each record corresponds to an AR, 
contains an M- or X-class flare as well as a CME associated with the flare,
and is tagged by whether the flare/CME produce an SEP event.
In this database, there are 31 records tagged by ``yes" indicating 
they are associated with SEP events while
there are 97 records tagged by ``no" indicating they are not associated with SEP events.
In our second database, referred to as the F\_S database,
each record corresponds to an AR, 
contains an M- or X-class flare,
and is tagged by whether the flare produces an SEP event 
regardless of whether or not the flare initiates a CME.
In this database, there are 40 records tagged by ``yes" indicating 
they are associated with SEP events while 
there are 700 records tagged by ``no" indicating 
they are not associated with SEP events.

\vspace{6pt}
\section{Methodology} 
\label{sec:methodology}
\subsection{Prediction Tasks}
As mentioned in Section \ref{sec:intro}, we aim to solve the following two binary prediction problems.
{\bf [FC\_S problem]}
Given a data sample $x_t$ at time point $t$ 
in an AR where the AR will produce an M- or X-class flare within the next $T$ hours of $t$ and the flare initiates a CME, we predict whether $x_t$ is positive or negative. Predicting $x_{t}$ to be positive means that the AR will produce an SEP event associated with the flare/CME. Predicting $x_{t}$ to be negative means that the AR will not produce an SEP event associated with the flare/CME.
{\bf [F\_S problem]} 
Given a data sample $x_t$ at time point $t$ in an AR where the AR will produce an M- or X-class flare within the next $T$ hours of $t$ regardless of whether or not the flare initiates a CME, 
we predict whether $x_t$ is positive or negative.
Predicting $x_{t}$ to be positive means that 
the AR will produce an SEP event associated with the flare. 
Predicting $x_{t}$ to be negative means that
the AR will not produce an SEP event associated with the flare.
For both of the two binary prediction problems, 
we consider $T$ ranging from 12 to 72 in 12-hour intervals 
as frequently considered in the literature 
\citep{Ahmed2013SoPh..283..157AFeatureSelection,BobraCME2016ApJ...821..127B,Inceoglu_2018CMESEP,Liu_2020CMEPrediction}.

\begin{figure*}
\centering
\gridline{\fig{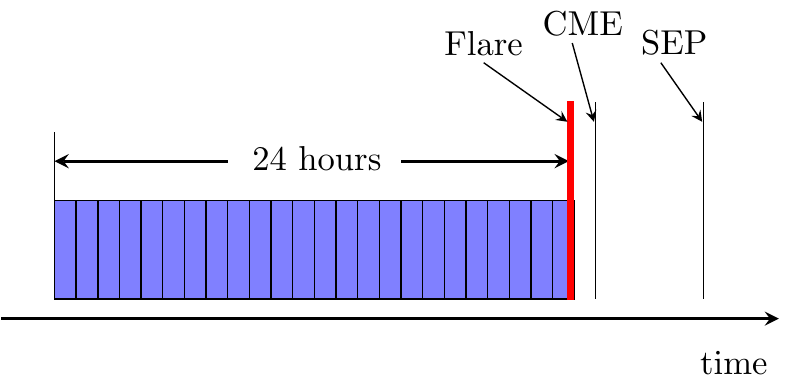}{0.49\textwidth}{\hspace{-0.5cm}\large{(a)}}
          \fig{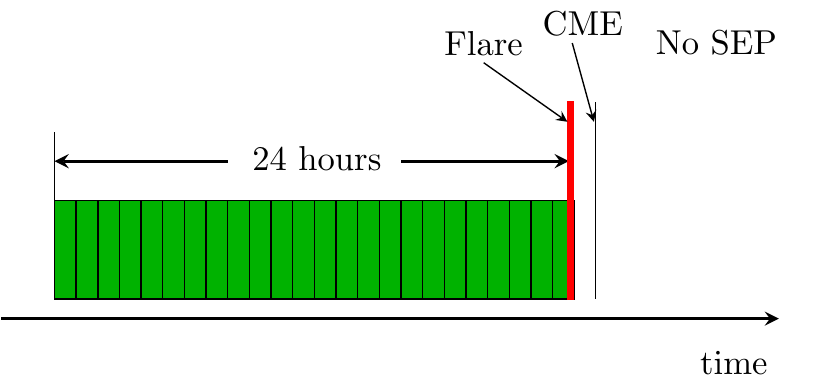}{0.49\textwidth}{\hspace{-0.5cm}\large{(b)}}}
          
\gridline{\fig{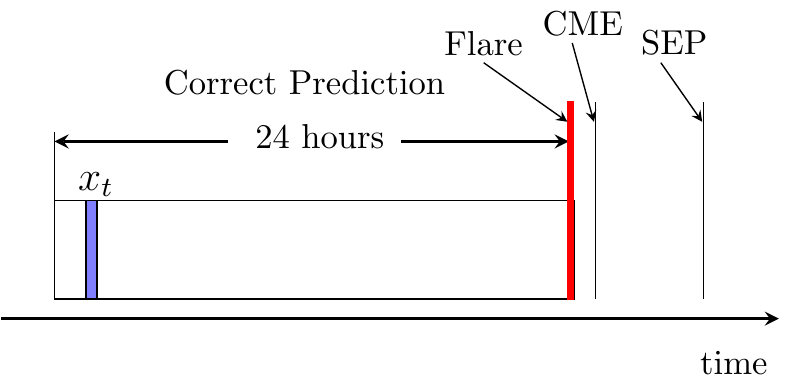}{0.49\textwidth}{\hspace{-0.5cm}\large{(c)}}
          \fig{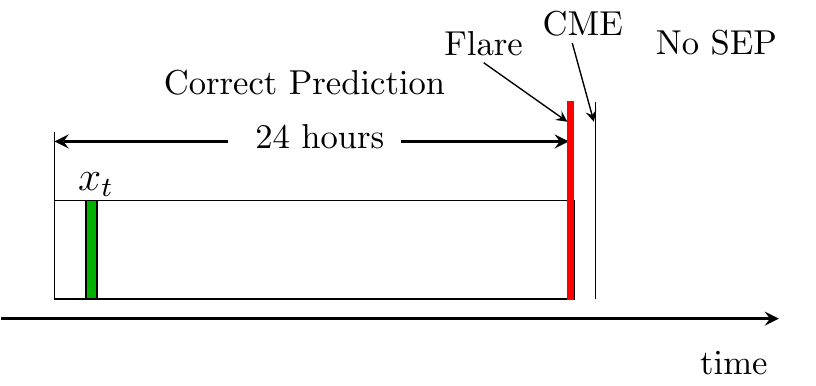}{0.49\textwidth}{\hspace{-0.5cm}\large{(d)}}}

\gridline{\fig{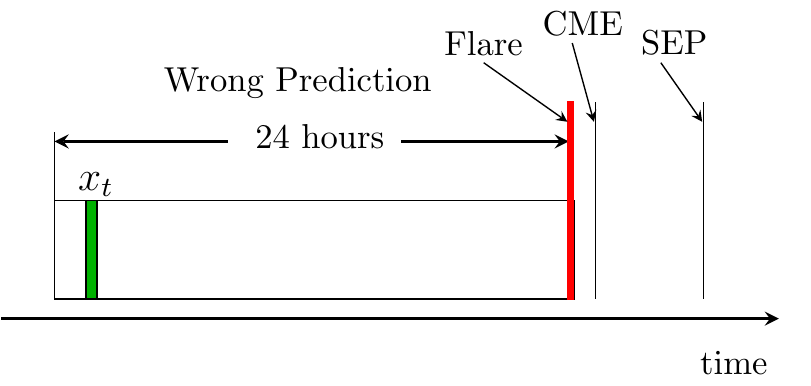}{0.49\textwidth}{\hspace{-0.5cm}\large{(e)}}
          \fig{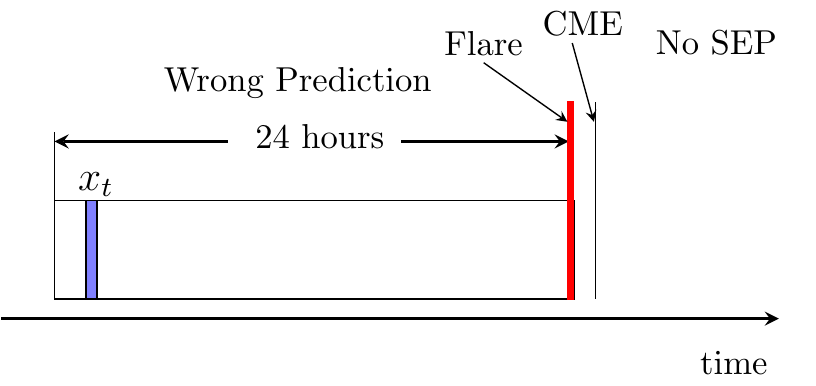}{0.49\textwidth}{\hspace{-0.5cm}\large{(f)}}}
\caption{Collecting and constructing positive and negative data samples 
on a flaring AR 
for the FC\_S problem where $T = 24$ hours
and making predictions based on the collected data samples. 
The data samples are collected at a cadence of 12 minutes. 
Each rectangular box corresponds to 1 hour and contains 5 data samples.
The red vertical line shows the peak time of 
an M- or X- class flare.
(a) The blue rectangular boxes contain data samples 
that are within the 24 hours prior to the peak time of 
an M- or X- class flare
that produces a CME and an SEP event; these blue data samples belong to the positive class.
(b) The green rectangular boxes contain data samples that are within the 24 hours 
prior to the peak time of 
an M- or X- class flare
that produces a CME but no SEP event; these green data samples belong to the negative class.
(c) Illustration of a correct prediction
for a test data sample $x_{t}$ that is predicted to be positive.
(d) Illustration of a correct prediction 
for a test data sample $x_{t}$ that is predicted to be negative.
(e) Illustration of a wrong prediction
for a test data sample $x_{t}$ that is predicted to be negative.
(f) Illustration of a wrong prediction
for a test data sample $x_{t}$ that is predicted to be positive.
}
\label{fig:sep_data_samples}
\end{figure*}

In solving the two
binary prediction problems, we first show how to
collect and 
construct positive and negative data samples 
used in our study.
Figure \ref{fig:sep_data_samples}(a) (Figure \ref{fig:sep_data_samples}(b), respectively) 
illustrates how to
construct positive (negative, respectively) data samples
for the FC\_S problem where $T = 24$ hours.
Refer to the FC\_S database described in
Section \ref{sec:data}, which indicates whether a flaring AR
that already produces an M- or X- class flare/CME
will initiate an SEP event associated with the flare/CME.
For the flaring AR, we collect data samples 
that are within the $T = 24$ hours prior to the peak time of the flare.
\begin{itemize}
\item 
If the flare/CME are associated with an SEP event,
the data samples belong to the positive class and are colored (labeled) by blue as shown in Figure \ref{fig:sep_data_samples}(a).
Thus, for each blue (positive) data sample, 
there is an M- or X- class flare that is
within the next 24 hours of the occurrence time of the data sample,
the flare initiates a CME and the flare/CME are associated with an SEP event.
\item
If the flare/CME are not associated with an SEP event,
the data samples belong to the negative class and are colored (labeled) by green as shown in Figure \ref{fig:sep_data_samples}(b).
Thus, for each green (negative) data sample, there is an M- or X- class flare that is
within the next 24 hours of the occurrence time of the data sample,
the flare initiates a CME but the flare/CME are not associated with an SEP event.
\end{itemize}
Constructing positive and negative data samples for the F\_S problem is done
similarly and its description is omitted.

\begin{deluxetable*}{lllcccccc}
\tablecaption{Numbers of Positive and Negative Data Samples 
Constructed for Different Hours for the FC\_S and F\_S Problems Respectively}
\label{tab:numpositivenegativef}
\tablewidth{0pt}
\tablehead{\colhead{}&{}&{}& \colhead{12 hr} & \colhead{24 hr} & \colhead{36 hr} & \colhead{48 hr} & \colhead{60 hr} &\colhead{72 hr}}
\startdata
\noalign{\smallskip}
FC\_S		&		& Positive & 994 & 2017 & 3055 & 4143 & 5221 & 6336\\
 & & Negative & 2952 & 5522 & 7864 & 9976 & 11687 & 13135\\
\noalign{\smallskip}
\noalign{\smallskip}
\hline
\noalign{\smallskip}
F\_S & & Positive & 1260 & 2561 &3863 & 5207 & 6517 & 7864 \\
 & & Negative & 19593 & 31534 & 40619 & 48189 & 54718 & 59821\\
\noalign{\smallskip}
\noalign{\smallskip}			
\enddata
\end{deluxetable*}
\vspace{-25pt}
Table \ref{tab:numpositivenegativef} shows the numbers of positive and negative data samples 
constructed for the FC\_S and F\_S problems respectively.
Consider the FC\_S problem. 
The positive and negative data samples 
are constructed based on the
31 records tagged by ``yes'' and 97 records tagged by ``no'' in the FC\_S database
described in Section \ref{sec:data}.
When $T$ = 24 hours and the cadence is 12 minutes,
one would expect the total number of positive data samples to be 
$24$ hr $\times$ 60 minutes/hr $\times$ (1/12 minutes) $\times$ 31 = 3720,
and the total number of negative data samples to be
$24$ hr $\times$ 60 minutes/hr $\times$ (1/12 minutes) $\times$ 97 = 11640. 
However,
the total number of positive (negative, respectively) data samples is 2017
(5522, respectively). 
This happens because we removed many data samples of low quality
as described in Section \ref{sec:data}.
If a gap occurs in the middle of a time series due to the removal,
we use a zero-padding strategy as done in \citet{Liu_2020CMEPrediction}
to create a synthetic data sample to fill the gap.
The synthetic data sample has zero values 
for all the 18 SHARP parameters.
The synthetic data sample is added after the normalization 
of the SHARP parameter values, and hence
the synthetic data sample does not affect the normalization procedure.

After explaining how to construct the positive and negative
data samples, we now show how to solve the binary prediction problems.
Consider again the FC\_S problem where $T = 24$ hours.
Here we want to predict whether a given test data sample
$x_{t}$ at time point $t$ is positive (blue) or negative (green)
given that there will be an M- or X- class flare
within the next 24 hours of $t$ and
the flare initiates a CME.
If there is an SEP event associated with the flare/CME, and we predict
$x_{t}$ to be positive (blue), then this is a correct prediction
as illustrated in Figure \ref{fig:sep_data_samples}(c).
If there is an SEP event associated with the flare/CME, but we predict
$x_{t}$ to be negative (green), then this is a wrong prediction
as illustrated in Figure \ref{fig:sep_data_samples}(e).
On the other hand, if there is no SEP event associated with the flare/CME,
and we predict $x_{t}$ to be negative (green), then this is a correct prediction
as illustrated in Figure \ref{fig:sep_data_samples}(d).
If there is no SEP event associated with the flare/CME,
but we predict $x_{t}$ to be positive (blue), then this is a wrong prediction
as illustrated in Figure \ref{fig:sep_data_samples}(f).
The F\_S problem is solved similarly.
In the following subsection we describe how to train 
our model and use the trained model to make predictions.

\newpage
\subsection{Prediction Method} \label{sec:predictionmethod}
We consider one of recurrent neural networks (RNNs) that is called 
long short-term memory \citep[LSTM;][]{LSTMHochreiter1997LongSM, Goodfellow_DeepLearningBookDBLP:books/daglib/0040158} 
to build our model. 
LSTM has shown good results in solar eruption prediction \citep{Liu_2019FlarePrediction,Liu_2020CMEPrediction}. 
We create a model using bidirectional LSTM (biLSTM). 
Generally, a bidirectional RNN \citep{BiRRNSchuster1997} functions by duplicating the initial recurrent layer in the network 
to obtain two layers so that one layer uses the input as is 
and the other duplicated layer uses the input in a reverse order. 
This design allows biLSTM to discover additional
patterns
that cannot be found by LSTM with only one recurrent layer \citep{LSTMbiLSTMComparison2019}.
In addition, the data used in our study is time series 
and biLSTM has shown an improvement over LSTM for general time series forecasting \citep{biLSTMEStockMarket2018,Kang2020TimeSP}.
As our experimental results show later, biLSTM also outperforms LSTM
in SEP prediction.

\begin{figure}
\centering
\includegraphics[]{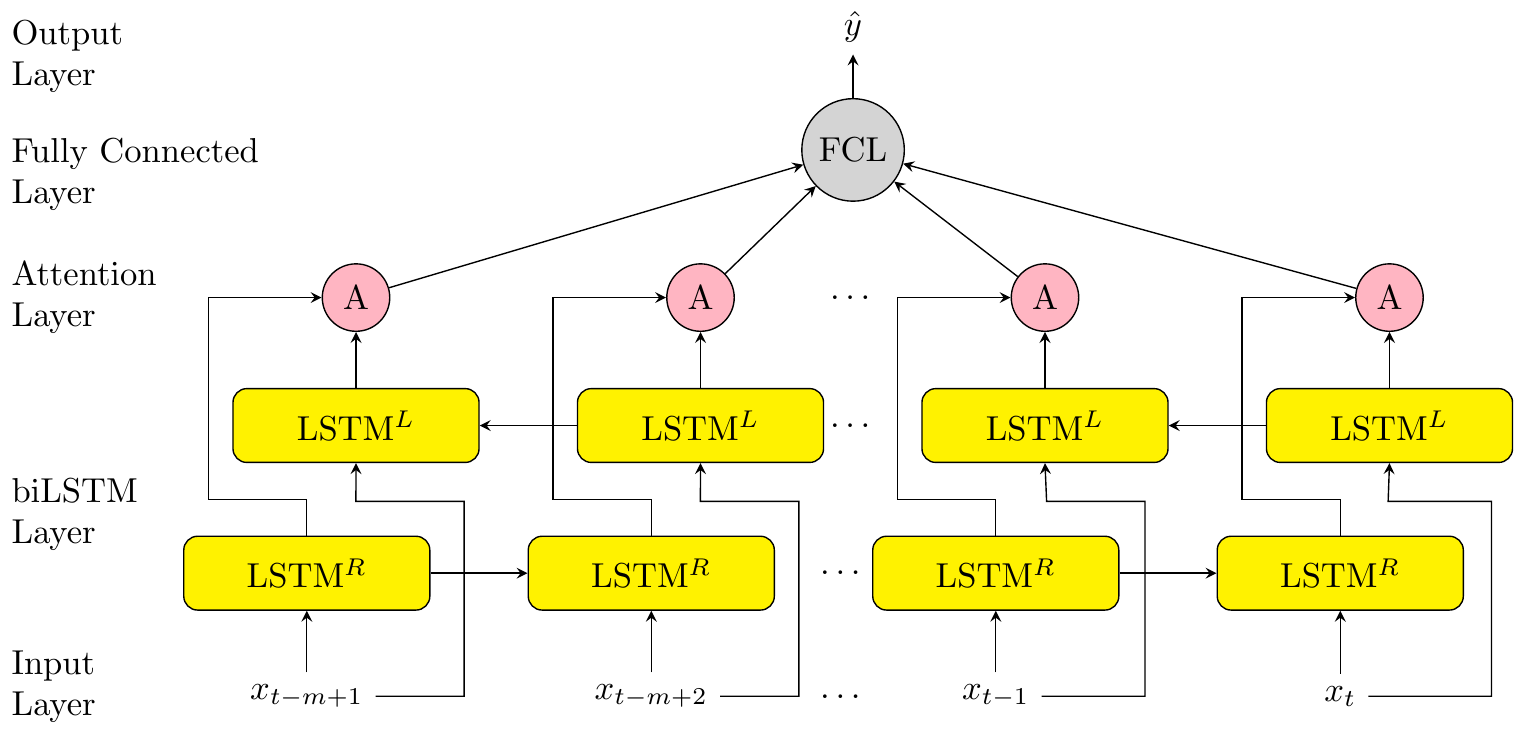}
\caption{Architecture of the proposed biLSTM network. 
Yellow boxes represent biLSTM cells.
These cells are connected to an attention layer (A) that contains $m$ neurons,
which are connected to a fully connected layer (FCL). 
(In the study presented here, $m$ is set to 10.)
During testing/prediction,
the input to the network is a test data sequence
with $m$ consecutive data samples $x_{t-m+1}$, $x_{t-m+2}$ \dots $x_{t-1}$, $x_{t}$
where $x_{t}$ is the test data sample at time point $t$.
The trained biLSTM network predicts the label (color)
of the test data sequence, more precisely the label (color) of $x_{t}$.
The output layer of the biLSTM network calculates a probability ($\hat{y}$) between 0 and 1.
If $\hat{y}$ is greater than or equal to a
threshold, which is set to 0.5, 
the biLSTM network outputs 1 and predicts $x_{t}$ to be positive,
i.e., predicts the label (color) of $x_{t}$ to be blue;
see Figure \ref{fig:sep_data_samples}.
Otherwise, the biLSTM network outputs 0 and 
predicts $x_{t}$ to be negative,
i.e., predicts the label (color) of $x_{t}$ to be green;
see Figure \ref{fig:sep_data_samples}.
\label{fig:biLSTM_arch}}
\end{figure}

Figure \ref{fig:biLSTM_arch} presents 
the architecture of our neural network, 
which accepts as input a data sequence with 
$m$ consecutive data samples.
(In the study presented here, $m$ is set to 10.)
The neural network consists of a biLSTM layer configured with 400 neurons.
In addition, the neural network contains
an attention layer motivated by \citet{Goodfellow_DeepLearningBookDBLP:books/daglib/0040158} 
to direct the network to focus on important information and characteristics of input data samples. 
The attention layer is designed to map and capture the alignment between the input and output 
by calculating a weighted sum for input data sequences. 
Specifically, the attention context vector 
for the output $\hat{y}_i$, denoted $\textbf{\textit{CV}}_{i}$, is calculated as follows:
\begin{equation}
\textbf{\textit{CV}}_i = \sum_{j=1}^{m}{\textbf{\textit{W}}_{i,j}\textbf{\textit{H}}_j},
\end{equation}
where $m$ is the input sequence length, 
${\textbf{\textit{H}}_j}$ is the hidden state corresponding to the input data sample $x_{j}$ and
\textbf{\textit{W}} contains weights applied to the hidden state.
\textbf{\textit{W}} is computed by a softmax function as follows:
\begin{equation}
\textbf{\textit{W}}_{i,j} = \frac{e^{S_{i,j}}}{\sum_{k=1}^{m}{e^{S_{i,k}}}}.
\end{equation}
Here $S_{i,j}$ is a score function calculated as follows:
\begin{equation}
S_{i,j} = \mathbfit{V}\times \mbox{tanh}(\mathbfit{W}^{'}(\mathbfit{S}_{i},\mathbfit{H}_{j})),
\end{equation}
where 
$\mbox{tanh}(\cdot)$ is the hyperbolic tangent function,
$\mathbfit{S}_{i}$ is the output state corresponding 
to the output $\hat{y}_{i}$,
\textbf{\textit{V}} and $\textbf{\textit{W}}^{'}$ are weight matrices learned by the neural network.
The attention layer passes its resulting vector to a fully connected layer.

\begin{figure}
 \includegraphics[width=2cm]{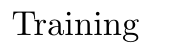}\\
 \gridline{
 \fig{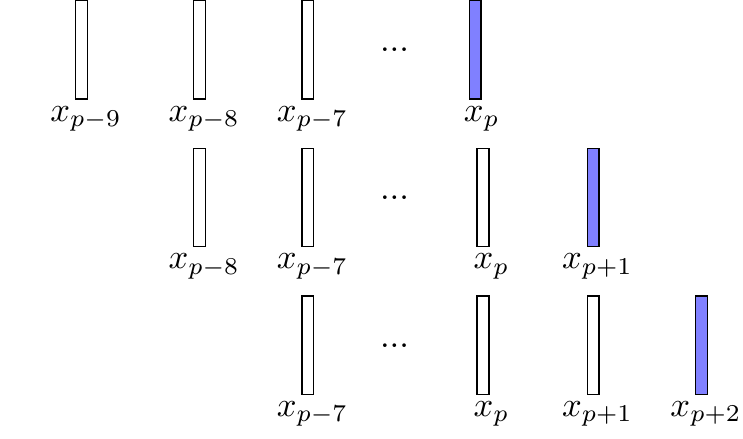}{0.49\textwidth}{\large{(a)}}
 \fig{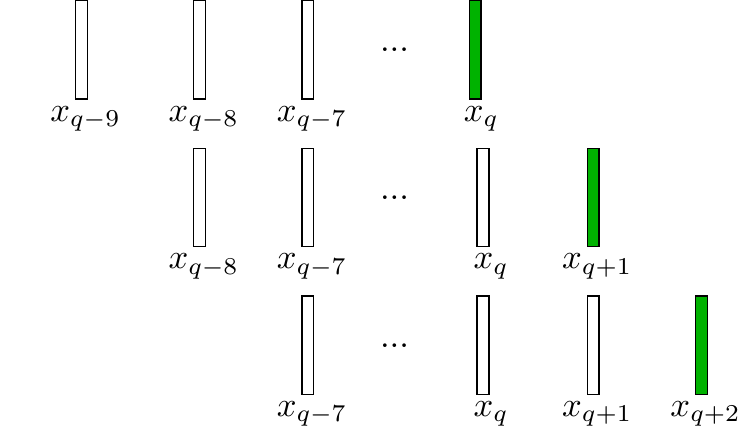}{0.49\textwidth}{\large{(b)}}
 }
 \includegraphics[width=2cm]{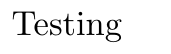}
 \gridline{
    \fig{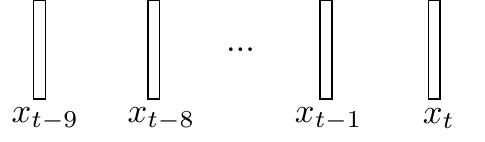}{0.32\textwidth}{\large{(c)}}
 }
 \caption{Example data sequences used to train and test our biLSTM network where each data sequence contains 10 consecutive data samples.
 (a) Three positive training data sequences taken from a flaring AR. 
 (b) Three negative training data sequences taken from a flaring AR.
 In (a) and (b),
 the label (color) of a training data sequence is
 defined to be the label (color) of the last data sample in the training data sequence while the labels (colors) of the other nine data samples in the training data sequence are ignored.
 (c) A test data sequence formed for predicting the label (color) of the last data sample $x_{t}$ in a flaring AR.
 }
 \label{fig:trainingtestingmodelfeed}
\end{figure}
During training, our biLSTM network
takes as input overlapping data sequences where each data sequence
contains $m = 10$ consecutive data samples.
The label (color) of a training data sequence is defined to be the label (color)
of the last (i.e., 10th) data sample in the data sequence
while the labels (colors) of the other nine data samples in the data sequence are ignored.
Thus, if the 10th data sample is positive (blue), then 
the training data sequence is positive;
if the 10th data sample is negative (green), then the training 
data sequence is negative.
We feed one training data sequence at a time to our biLSTM network
when training the model. 
Figure \ref{fig:trainingtestingmodelfeed}(a) illustrates
three positive data sequences used to train our biLSTM model.
Figure \ref{fig:trainingtestingmodelfeed}(b) illustrates
three negative data sequences used to train our biLSTM model.

The loss function used in our biLSTM model is the weighted binary cross-entropy (WBCE) \citep{Goodfellow_DeepLearningBookDBLP:books/daglib/0040158,Liu_2020CMEPrediction}. 
Let $N$ denote the total number of data sequences each having $m$ consecutive data samples in the training set.
Let $w_0$ denote the weight for the positive class 
(i.e., minority class)
and let $w_1$ denote the weight for the negative class
(i.e., majority class).
The weights are calculated based on the ratio of majority and minority class sizes 
with more weight assigned to the minority class.
Let $y_i$ denote the observed probability
of the $i$th data sequence;
$y_i$ is 1 if the $i$th data sequence is positive 
and 0 if the $i$th data sequence is negative.
Let $\hat{y}_i$ denote the predicted probability 
of the $i$th data sequence.
The WBCE, calculated as follows, is suitable for imbalanced datasets such as those tackled here where
the negative class has more data samples than the positive class;
see Table \ref{tab:numpositivenegativef}.

\begin{equation}
 \mbox{WBCE} = \sum^N_{i=1} w_{0}y_{i}\log(\hat{y}_i) + w_1(1 - y_i)\log(1 - \hat{y}_i).
\end{equation}

We configure the network to use a fraction (1/10) of the training set 
as the internal validation subset.
We employ progressive learning with early stopping and
adopt the strategy of saving the highest performing model during the iterative learning process.
The performance of a model is measured by the WBCE on the internal validation subset 
where the smaller the WBCE is, the better performance the model has.
In each iteration, the process checks the performance of the models in the current and previous iterations 
to decide which model to use for the next iteration. 
If the model in the current iteration has better performance, 
the process copies its weights as starting weights for the next iteration; 
otherwise, it copies the weights of the model in the previous iteration
as starting weights for the next iteration.
This progressive process improves the weights of the network's hidden layers and as a result 
the overall performance of the network is also improved. 
In addition, during the iterations, if the performance of the network degrades, 
the process stops and selects the highest performing model it identifies within the iterations.

During testing/prediction, 
we are given a test data sample $x_{t}$
and our biLSTM model will predict the label (color) of $x_{t}$,
i.e., predict whether $x_{t}$ is positive or negative.
We pack the $m-1$ data samples preceding $x_{t}$, namely
$x_{t-m+1}$, $x_{t-m+2}$, $\ldots$, $x_{t-1}$, along with $x_{t}$
into a test data sequence with $m$ consecutive data samples and feed this test data sequence 
to our biLSTM model
as shown in the input layer in Figure \ref{fig:biLSTM_arch}.
Figure \ref{fig:trainingtestingmodelfeed}(c) illustrates
a test data sequence where $m$ is 10.
The output layer of our biLSTM model 
calculates a probability between 0 and 1 for the test data sequence. 
We compare the probability with a threshold,
which is set to 0.5.
If the probability is greater than or equal to the threshold,
our biLSTM model outputs 1 indicating the test data sequence, more precisely the test data sample $x_{t}$, is positive;
otherwise our model outputs 0 indicating the test data sequence, more precisely $x_{t}$, is negative.

\section{Results} 
\label{sec:result}
\subsection{Performance Metrics and Experiment Setup}
\label{sec:performancemeterics}
We conducted a series of experiments to evaluate the performance of the proposed method and
compare it with related machine learning methods.
For the data sample $x_{t}$ at time point $t$, we define:
\begin{itemize}
\item 
$x_t$ to be true positive (TP) if our model (network) predicts that $x_t$ is positive and $x_t$ is indeed positive, 
i.e., an SEP event will be produced with respect to $x_{t}$;
\item 
$x_t$ to be false positive (FP) if our model predicts that $x_t$ is positive while $x_t$ is actually negative, 
i.e., no SEP event will be produced with respect to $x_{t}$;
\item
$x_t$ to be true negative (TN) if our model predicts $x_t$ is negative and $x_t$ is indeed negative;
\item
$x_t$ to be false negative (FN) if our model predicts $x_t$ is negative while $x_t$ is actually positive.
\end{itemize}
We also use TP (FP, TN, FN, respectively) to denote 
the total number of true positives (false positives, true negatives, false negatives, respectively) 
produced by a method.

The following performance metrics are used in our study:
\begin{equation}
	\text{Recall} = \frac{ \mathrm{TP}}{\mathrm{TP + FN}},
\end{equation}

	\begin{equation}
	\text{Precision} = \frac{ \mathrm{TP}}{\mathrm {TP + FP}},
\end{equation}
	\begin{equation}
		\text{Balanced Accuracy (BACC)} = \frac{1}{2} \left( \frac{ \mathrm{TP}}{\mathrm{TP+FN}} + \frac{ \mathrm{TN}}{\mathrm{TN+FP}}\right),
	\end{equation}

	\begin{equation}
		\text{Heidke Skill Score (HSS)}	= \frac{\text{2} \times (\mathrm{TP} \times \mathrm{TN - FP} \times \mathrm{FN})}{(\mathrm{TP + FN})\times(\mathrm{FN+TN)} + (\mathrm{TP + FP}) \times (\mathrm{FP+TN})},
	\end{equation}

	\begin{equation}
		\text{True Skill Statistics (TSS)} = \frac{ \mathrm{TP}}{\mathrm{TP+FN}} - \frac{\mathrm{FP}}{\mathrm{FP+TN}}.
	\end{equation}

BACC \citep{BACC_Paper_2009} is an accuracy measure mainly for 
imbalanced datasets.
HSS \citep{HSSHeidke1926} and TSS~\citep{Bloomfield:TSS:Recommend:2012ApJ...747L..41B}
are commonly used for flare, CME and SEP predictions
\citep{Bloomfield:TSS:Recommend:2012ApJ...747L..41B,ForecastingFlaresPredictors2018SoPh..293...28F,Inceoglu_2018CMESEP,Liu_2019FlarePrediction,Liu_2020CMEPrediction}. 
HSS ranges from $-\infty$ to $+1$. 
The higher HSS value a method has, the better performance the method achieves. 
TSS ranges from $-1$ to $+1$.
Like HSS, the higher TSS value a method has, the better performance the method achieves. 
In addition, we use the weighted area under the curve (WAUC)
\citep{WAUCBekkar2013EvaluationMF} 
in our study.
The area under the curve (AUC) in a receiver operating characteristic (ROC) curve \citep[][]{AUC2004WtFor..19.1106M}
indicates how well a method is capable of
distinguishing between two classes in binary prediction with the ideal value of one.
When calculating the AUC, we do not distinguish between the accuracy on
the minority class (positive class) and the accuracy on the majority class (negative class).
In contrast, when calculating the WAUC, which is an extension of the AUC and mainly for imbalanced datasets
like those tackled here, 
the accuracy on the minority class has a larger contribution 
to the overall performance of a model than the accuracy on the majority class.
As a consequence, we assign more weight to the accuracy on the minority class where
the weight is defined to be the ratio of the sizes of the minority and
majority classes.
All the metrics mentioned above are calculated using the confusion matrices 
obtained from the cross-validation (CV) scheme. 
With CV, we train a model using a subset of data, called the training set, 
and test the model using another subset of data, called the test set,
where the training set and test set are disjoint.
We consider six years, namely 2011-2015 and 2017, as mentioned in Section \ref{sec:data}.
Data samples from each year in turn are used for testing in a run and
data samples from all the other five years together are used for training 
in the run.
There are six years, and hence there are six runs in total. 
For each performance metric, the mean and standard deviation 
over the six runs are calculated and recorded.

\subsection{Parameter Ranking and Selection}
\label{sec:featureselection}
We first assessed the importance of the 18 SHARP parameters described in 
Section \ref{sec:data} to understand which parameters are the most important ones with
the greatest predictive power by utilizing a parameter ranking method, called Stability Selection
\citep{StabilityFeatureSelection2010}. 
This method is based on the 
LASSO (Least Absolute Shrinkage and Selection Operator) algorithm \citep{LASSO}.
Table \ref{tab:rankingfandfc} presents
the rankings
of the parameters with respect to $T$ = 12, 24, 36, 48, 60 and 72 for the FC\_S and F\_S problems respectively. 
The parameter ranked first is the most important one 
while the parameter ranked 18th is the least important one.
ABSNJZH is ranked consistently high for the FC\_S problem
while SAVNCPP and TOTUSJH are ranked high for the F\_S problem.
AREA\_ACR, TOTUSJZ and USFLUX are ranked consistently low for both of the FC\_S and F\_S problems.

\begin{deluxetable*}{lccccccccccccccccc}
	\tablecaption{Importance Rankings of the 18 SHARP Parameters Used in Our Study for the FC\_S and F\_S Problems Respectively
	\label{tab:rankingfandfc}}
	\tablewidth{\columnwidth}
	\tablehead{
		\multicolumn{1}{l}{\vspace{-0.2cm} SHARP}& \multicolumn{2}{c}{\raggedbottom 12 hr} & \colhead{} & \multicolumn{2}{c}{\raggedbottom 24 hr} &\colhead{} & \multicolumn{2}{c}{\raggedbottom 36 hr} &\colhead{} & \multicolumn{2}{c}{\raggedbottom 48 hr} &\colhead{} & \multicolumn{2}{c}{\raggedbottom 60 hr} & \colhead{} &\multicolumn{2}{c}{\raggedbottom 72 hr}\\
\multicolumn{1}{l}{Keyword} & \colhead{FC\_S} & \colhead{F\_S} & \colhead{} & \colhead{FC\_S} & \colhead{F\_S} & \colhead{} & \colhead{FC\_S} & \colhead{F\_S} & \colhead{} & \colhead{FC\_S} & \colhead{F\_S} & \colhead{} & \colhead{FC\_S} & \colhead{F\_S} & \colhead{ } & \colhead{FC\_S} & \colhead{F\_S} 
	}
	\setlength{\tabcolsep}{2.5pt}
	\startdata
ABSNJZH & 3 & 3 & & 1 & 4 & & 1 & 10 & & 1 & 10 & & 1 & 2 & & 5 & 1\\
AREA\_ACR & 17 & 16 & & 17 & 16 & & 17 & 16 & & 17 & 16 & & 16 & 16 & & 16 & 16\\
MEANALP & 13 & 15 & & 3 & 15 & & 3 & 15 & & 3 & 15 & & 2 & 15 & & 6 & 15\\
MEANGAM & 4 & 14 & & 4 & 14 & & 4 & 14 & & 4 & 14 & & 4 & 14 & & 8 & 14\\
MEANGBH & 5 & 13 & & 5 & 13 & & 5 & 13 & & 5 & 13 & & 14 & 13 & & 14 & 7\\
MEANGBT & 6 & 12 & & 6 & 12 & & 6 & 12 & & 6 & 12 & & 13 & 12 & & 13 & 13\\
MEANGBZ & 7 & 11 & & 7 & 3 & & 7 & 4 & & 7 & 4 & & 12 & 11 & & 4 & 4\\
MEANJZD & 8 & 10 & & 8 & 11 & & 8 & 11 & & 8 & 11 & & 11 & 10 & & 12 & 12\\
MEANJZH & 2 & 9 & & 2 & 10 & & 2 & 5 & & 2 & 5 & & 10 & 9 & & 11 & 10\\
MEANPOT & 10 & 8 & & 10 & 9 & & 10 & 9 & & 10 & 9 & & 9 & 8 & & 1 & 11\\
MEANSHR & 11 & 7 & & 11 & 8 & & 11 & 8 & & 11 & 8 & & 8 & 7 & & 10 & 3\\
R\_VALUE & 12 & 6 & & 12 & 7 & & 12 & 1 & & 12 & 3 & & 7 & 6 & & 2 & 2\\
SAVNCPP & 1 & 2 & & 13 & 2 & & 13 & 3 & & 13 & 1 & & 6 & 1 & & 3 & 5\\
SHRGT45 & 14 & 5 & & 14 & 6 & & 14 & 7 & & 14 & 7 & & 5 & 5 & & 9 & 9\\
TOTPOT & 15 & 4 & & 15 & 5 & & 15 & 6 & & 15 & 6 & & 15 & 4 & & 15 & 8\\
TOTUSJH & 9 & 1 & & 9 & 1 & & 9 & 2 & & 9 & 2 & & 3 & 3 & & 7 & 6\\
TOTUSJZ & 16 & 17 & & 16 & 17 & & 16 & 17 & & 16 & 17 & & 17 & 17 & & 17 & 17\\
USFLUX & 18 & 18 & & 18 & 18 & & 18 & 18 & & 18 & 18 & & 18 & 18 & & 18 & 18\\
	\enddata
\end{deluxetable*}
\vspace{-24pt}
\begin{figure*}
	\centering
 \gridline{
 \fig{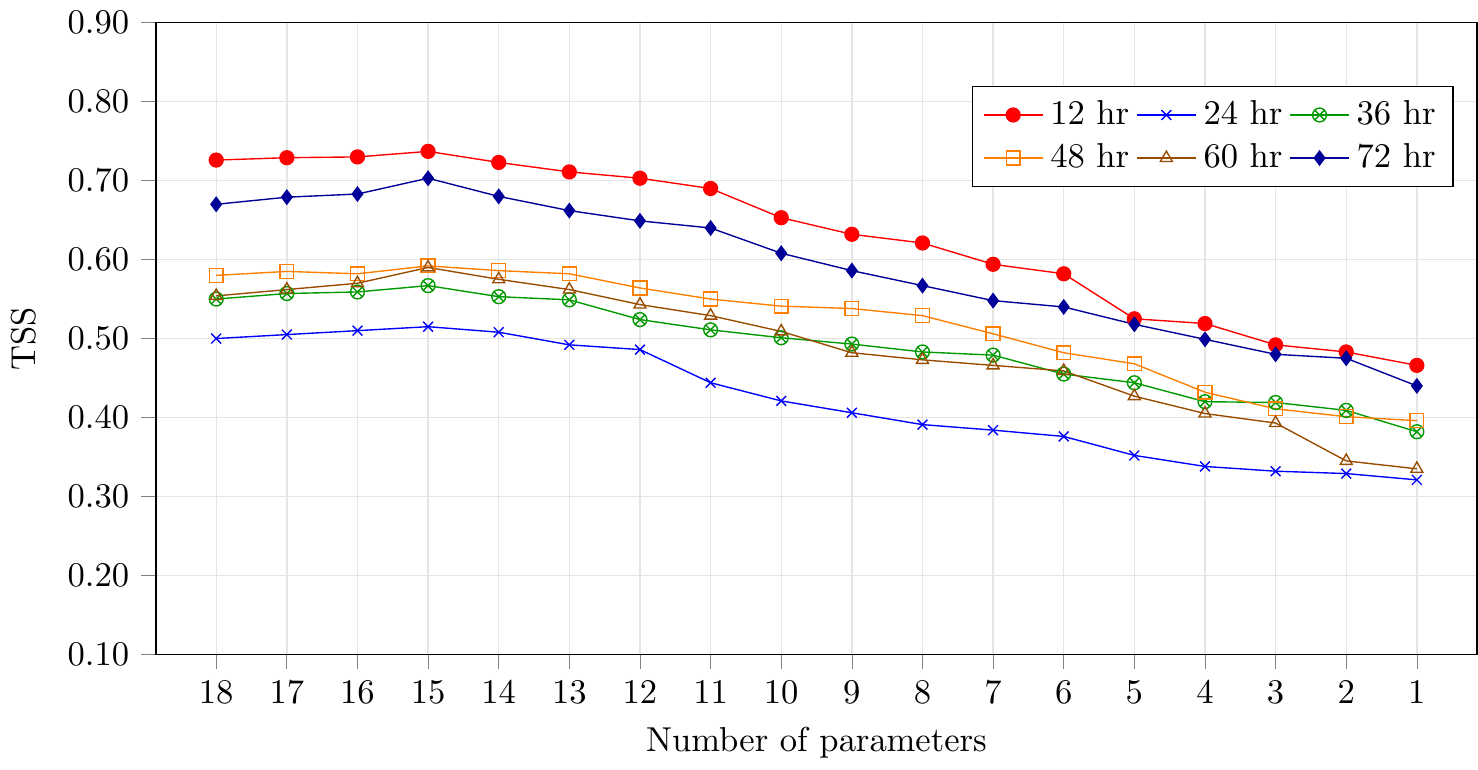}{0.90\textwidth}{\hspace{1.3cm}\large{(a)}}
 }
  \gridline{
 \fig{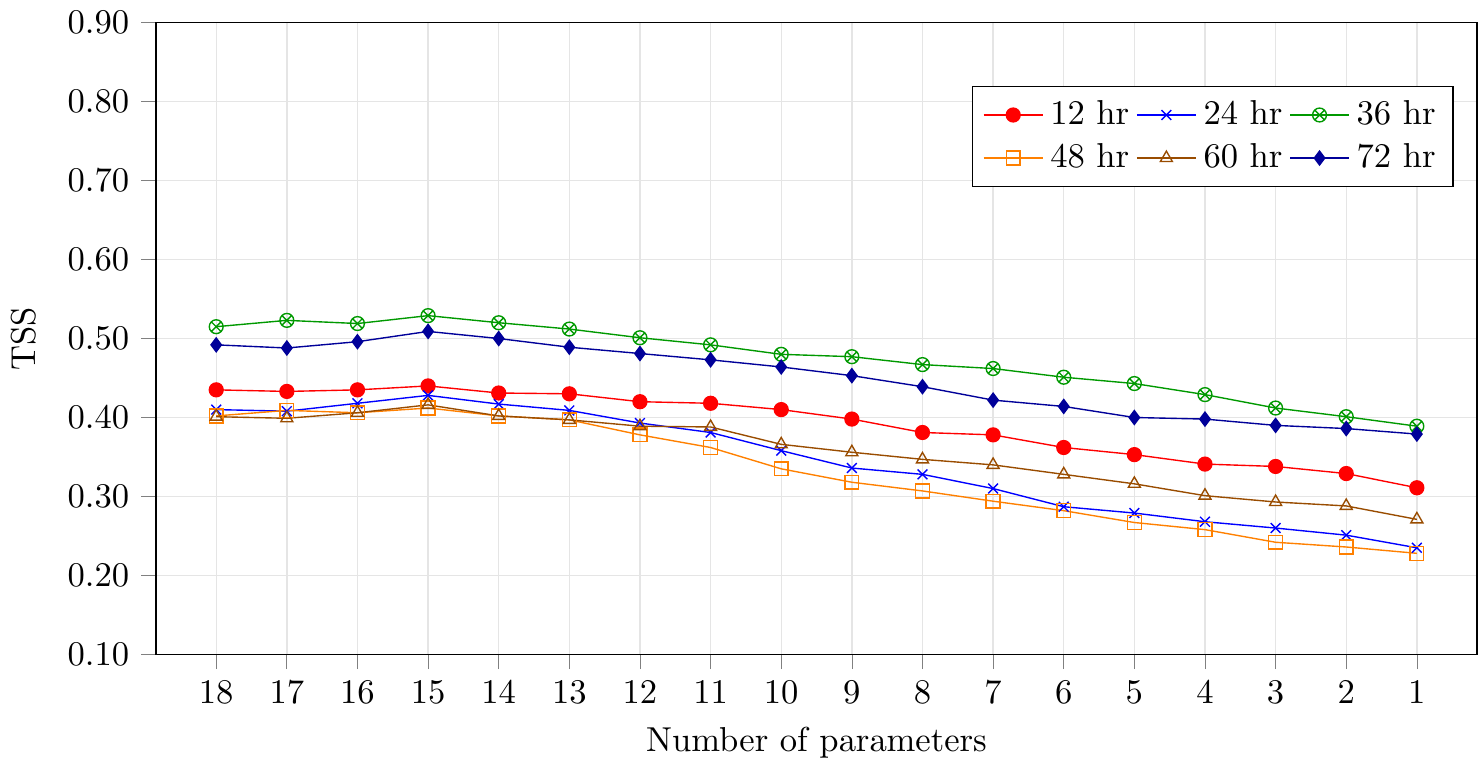}{0.90\textwidth}{\hspace{1.3cm}\large{(b)}} 
 }
	\caption{Parameter selection results for (a) the FC\_S problem, and (b) the F\_S problem.}
	\label{fig:f_recursiveselection}
\end{figure*}
We then used the recursive parameter elimination algorithm \citep{FeatureEngineeringSelection} 
in combination with our biLSTM model
to select a set of parameters that achieves the
best performance where the performance is measured by TSS.
The parameter elimination algorithm is an interactive procedure.
It selects parameters by recursively considering smaller and smaller sets of parameters where
the least important parameters are successively pruned from the current set of parameters. 
Figure \ref{fig:f_recursiveselection} 
presents the parameter selection results for the FC\_S and F\_S problems respectively.
It can be seen from the figure that using the top 15 most important parameters 
achieves the best performance
for both of the FC\_S and F\_S problems.
When using the top $k$, $1 \leq k \leq 14$, most important parameters, 
the less parameters we use,
the worse performance our model achieves.
Using the top-ranked, most important parameter alone would yield a lower TSS than
using all the top 15 most important parameters together. 
In subsequent experiments, we used the top 15 most important parameters for our biLSTM model.
That is, we removed the three least important parameters
AREA\_ACR, TOTUSJZ and USFLUX from data samples
and each data sample contained only the
top 15 most important SHARP parameters.
\subsection{Performance Comparison} 
\label{sec:performanceresultandcomparison}
Next, we compared our biLSTM network with four related machine learning methods, including
multilayer perceptrons (MLP),
support vector machines (SVM), random forests (RF) and long short-term memory (LSTM) \citep{Liu_2019FlarePrediction}.
These four methods are commonly used to
predict solar flares, CMEs and SEPs \citep{BobraCME2016ApJ...821..127B,Liu..Wang..Solar..2017ApJ...843..104L,ForecastingFlaresPredictors2018SoPh..293...28F,Inceoglu_2018CMESEP, CMH-2019,Liu_2019FlarePrediction,Liu_2020CMEPrediction,WCT-2020, MLaaS_RAA_Abduallah_2021}.

MLP \citep{ThePerceptron..book..1958, MLPARIASDELCAMPO2021115147} is a feed-forward artificial neural network \citep{ANN:DBLP:conf/ann/1995} that consists of an input layer,
an output layer, and one or more hidden layers. 
The number of hidden layers is set to 
3 with 200 neurons in each hidden layer.
SVM \citep{SVM_Cristianini2008} is trained with the Radial Basis Function (RBF) kernel and the cache size is set to 20000 to speed the training process.
RF \citep{RandomForestCART1984} is an ensemble algorithm 
that has two hyperparameters for performance tuning: $m$ 
(the number of SHARP magnetic parameters randomly selected and used to split a node 
in a tree of the forest) 
and $n$ (the number of trees to grow). 
We set $m$ to 2 and $n$ to 500.
The implementation of LSTM follows that described in \citet{Liu_2020CMEPrediction}.
The hyperparameters not specified here are set to their default values provided by
the scikit-learn library in Python \citep{scikit-learn}. 

As done in biLSTM, we used the recursive parameter elimination algorithm \citep{FeatureEngineeringSelection} to identify and select the best parameters for the four related machine learning methods
based on the importance rankings of the 18 SHARP parameters in 
Table \ref{tab:rankingfandfc} 
for the FC\_S and F\_S problems respectively.
Our experiments showed that, like biLSTM, using the top 15 most important parameters
achieved the best performance for the four related machine learning methods.
Consequently, we used the top 15 most important parameters 
for the four machine learning methods in the experimental study.

\begin{figure*}
\includegraphics[width=1\columnwidth]{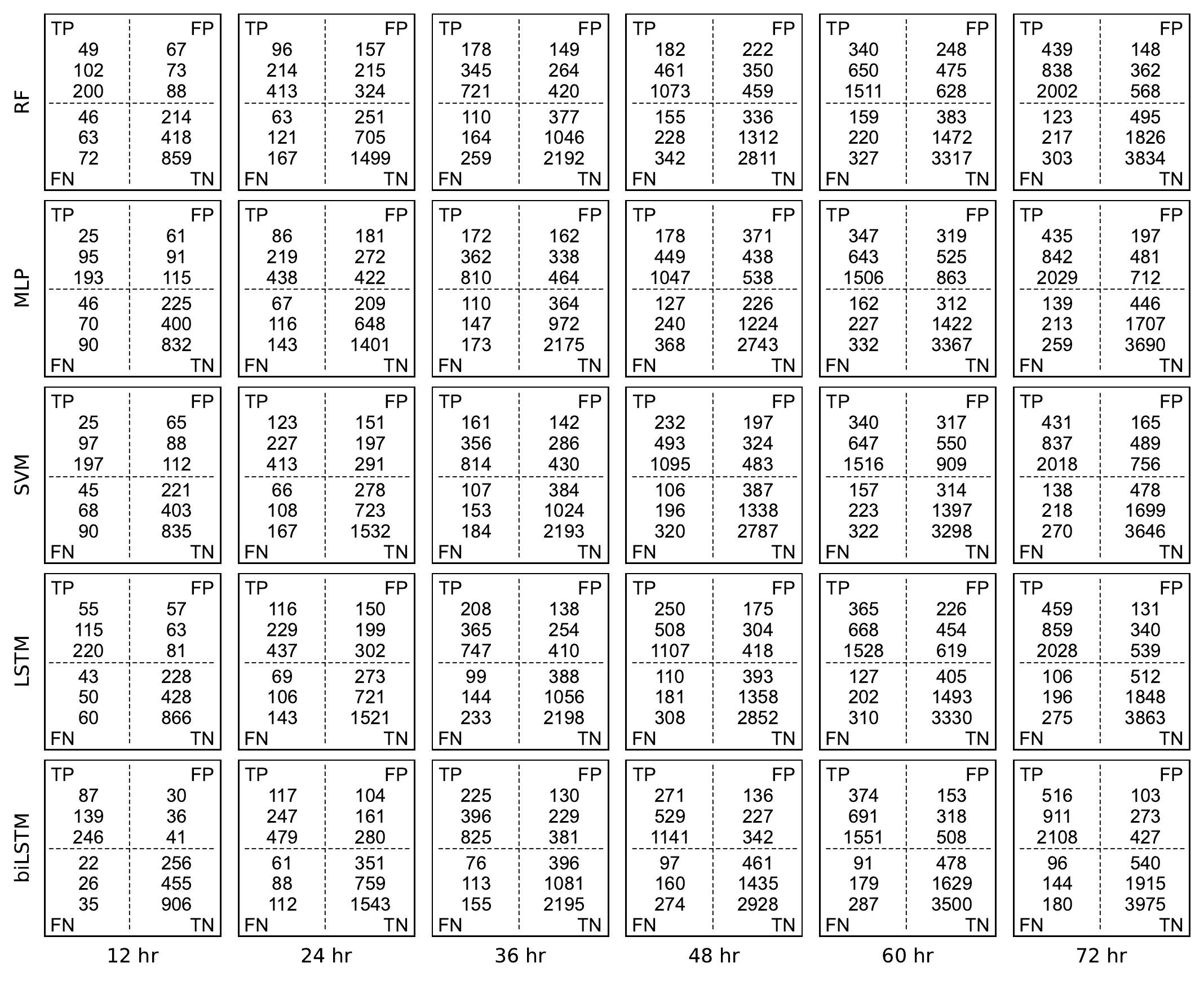}
\caption{Confusion matrices of RF, MLP, SVM, LSTM and biLSTM for the FC\_S problem. 
For each $T$, $T$ = 12, 24, 36, 48, 60, 72, and each machine learning method,
the figure shows the minimum, average, maximum
(displayed from top to bottom)
TP, FN, TN, FP respectively
from the six runs based on our cross-validation scheme.
\label{fig:confusionmatrixfc}}
\end{figure*}

\begin{figure*}[ht!]
 \includegraphics[width=1\columnwidth]{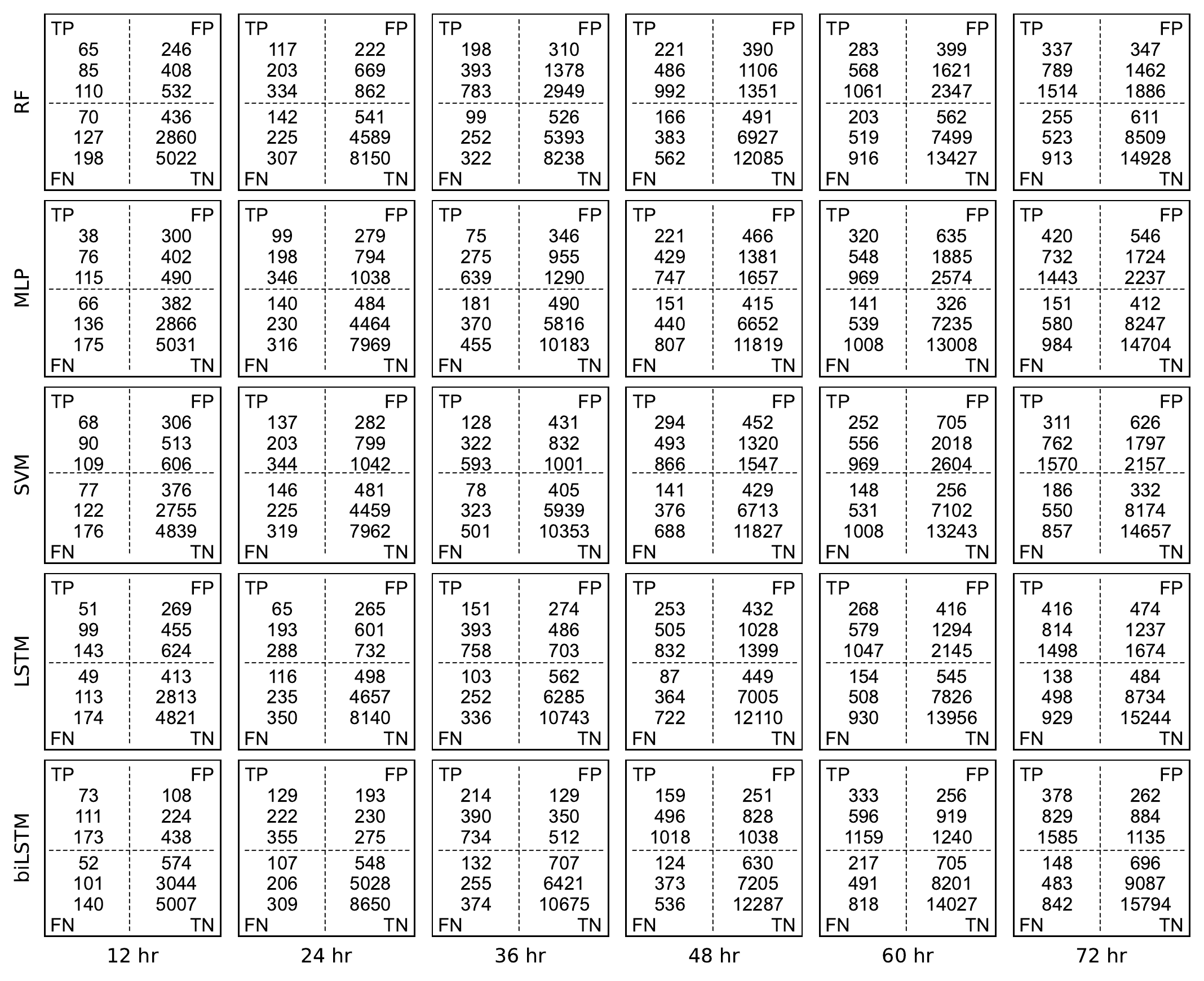}
\caption{Confusion matrices of RF, MLP, SVM, LSTM and biLSTM for the F\_S problem. 
For each $T$, $T$ = 12, 24, 36, 48, 60, 72, and each machine learning method,
the figure shows the minimum, average, maximum 
(displayed from top to bottom)
TP, FN, TN, FP respectively
from the six runs based on our cross-validation scheme.
\label{fig:confusionmatrixf}}
\end{figure*}

Figures \ref{fig:confusionmatrixfc} and \ref{fig:confusionmatrixf}
present the confusion matrices of the five machine learning methods (RF, MLP, SVM, LSTM, biLSTM) 
for the FC\_S and F\_S problems respectively. 
For each $T$, $T$ = 12, 24, 36, 48, 60, 72, and each machine learning method,
the figures show the minimum, average, maximum 
(displayed from top to bottom)
TP, FN, TN, FP respectively
from the six runs based on our cross-validation scheme.
For example, refer to $T$ = 12 and biLSTM in Figure \ref{fig:confusionmatrixfc}.
The minimum (maximum, respectively) TP 
obtained by biLSTM 
from the six runs
is 87 (246, respectively); 
the average TP over the six runs is 139.
It can be seen from Figures \ref{fig:confusionmatrixfc} and \ref{fig:confusionmatrixf} that 
the average TN values are much larger than the average FP values
for both of the FC\_S and F\_S problems. 
This happens because there are many negative training data samples
in our datasets (see Table \ref{tab:numpositivenegativef}).
As a consequence, the machine learning methods gain sufficient knowledge about the negative data samples and hence
can detect them relatively easily.
For the FC\_S problem, 
the average TP values (TN values, respectively) are consistently larger than the average FN values (FP values, respectively),
indicating that the machine learning methods can solve the
FC\_S problem reasonably well.
For the F\_S problem, 
the average TP values are close to, or even smaller than, 
the average FN values in many cases, 
suggesting that
the machine learning methods have difficulty in
detecting positive data samples.
This is understandable given that there are much fewer
positive training data samples than negative training data samples
for the F\_S problem (see Table \ref{tab:numpositivenegativef}).

Tables \ref{tab:seppredictionresultsfc} and \ref{tab:seppredictionresultsf} 
compare the performance of the five machine learning methods 
for the FC\_S and F\_S problems respectively. 
The tables present the mean performance metric values
averaged over the six runs based on our cross-validation scheme
with standard deviations enclosed in parentheses.
Best average metric values are highlighted in boldface.
It can be seen from Tables \ref{tab:seppredictionresultsfc} and \ref{tab:seppredictionresultsf} that our biLSTM network
outperforms the four related machine learning methods
in terms of BACC, HSS, TSS and WAUC.
Furthermore, the five machine learning methods generally perform better in solving
the FC\_S problem than in solving the F\_S problem.
This result indicates that one can predict SEP events more accurately
when active regions will produce both flares and associated CMEs.
Using flare information alone to predict SEP events is harder and would produce less reliable prediction results.
\begin{deluxetable*}{llcccccc}[ht!]
	\tablecaption{Performance Comparison of RF, MLP, SVM, LSTM and biLSTM 
	Based on Our Cross-Validation Scheme 
	for the FC\_S Problem
	\label{tab:seppredictionresultsfc}}
\tablewidth{0pt}
	\tablehead{
		\colhead{}&{}& \colhead{12 hr} & \colhead{24 hr} & \colhead{36 hr} &
		\colhead{48 hr} & \colhead{60 hr} &\colhead{72 hr}
	}
	\startdata
Recall  &  RF & 0.593 (0.106)  & 0.617 (0.120)  & 0.660 (0.078)  & 0.632 (0.134)  & 0.721 (0.082)  & 0.767 (0.095)  \\
 &  MLP & 0.542 (0.175)  & 0.617 (0.149)  & 0.679 (0.110)  & 0.618 (0.143)  & 0.711 (0.088)  & 0.766 (0.093)  \\
 &  SVM & 0.543 (0.175)  & 0.666 (0.088)  & 0.663 (0.125)  & 0.686 (0.123)  & 0.714 (0.090)  & 0.761 (0.093)  \\
 &  LSTM & 0.658 (0.113)  & 0.663 (0.099)  & 0.699 (0.064)  & 0.714 (0.100)  & 0.745 (0.071)  & 0.788 (0.089)  \\
 &  biLSTM & \textbf{0.825} (0.058)  & \textbf{0.710} (0.112)  & \textbf{0.758} (0.064)  & \textbf{0.748} (0.086)  & \textbf{0.777} (0.077)  & \textbf{0.843} (0.058)  \\
\hline
Precision  &  RF & 0.554 (0.113)  & 0.483 (0.151)  & 0.556 (0.163)  & 0.532 (0.165)  & 0.548 (0.155)  & 0.661 (0.171)  \\
 &  MLP & 0.478 (0.153)  & 0.432 (0.166)  & 0.501 (0.155)  & 0.471 (0.161)  & 0.524 (0.164)  & 0.605 (0.161)  \\
 &  SVM & 0.486 (0.168)  & 0.520 (0.132)  & 0.543 (0.158)  & 0.574 (0.163)  & 0.514 (0.176)  & 0.603 (0.175)  \\
 &  LSTM & 0.609 (0.102)  & 0.516 (0.142)  & 0.581 (0.158)  & 0.599 (0.155)  & 0.569 (0.153)  & 0.680 (0.166)  \\
 &  biLSTM & \textbf{0.777} (0.061)  & \textbf{0.587} (0.148)  & \textbf{0.619} (0.149)  & \textbf{0.669} (0.155)  & \textbf{0.656} (0.160)  & \textbf{0.739} (0.133)  \\
\hline
BACC  &  RF & 0.707 (0.045)  & 0.672 (0.047)  & 0.719 (0.036)  & 0.688 (0.051)  & 0.713 (0.030)  & 0.790 (0.043)  \\
 &  MLP & 0.663 (0.061)  & 0.641 (0.062)  & 0.690 (0.049)  & 0.641 (0.051)  & 0.691 (0.046)  & 0.752 (0.023)  \\
 &  SVM & 0.668 (0.070)  & 0.706 (0.022)  & 0.709 (0.045)  & 0.730 (0.032)  & 0.687 (0.053)  & 0.753 (0.029)  \\
 &  LSTM & 0.751 (0.047)  & 0.704 (0.035)  & 0.744 (0.029)  & 0.750 (0.031)  & 0.734 (0.025)  & 0.807 (0.041)  \\
 &  biLSTM & \textbf{0.868} (0.029)  & \textbf{0.757} (0.040)  & \textbf{0.784} (0.029)  & \textbf{0.796} (0.037)  & \textbf{0.795} (0.032)  & \textbf{0.852} (0.021)  \\
\hline
HSS  &  RF & 0.404 (0.091)  & 0.315 (0.100)  & 0.406 (0.088)  & 0.354 (0.100)  & 0.386 (0.063)  & 0.545 (0.099)  \\
 &  MLP & 0.314 (0.119)  & 0.249 (0.123)  & 0.343 (0.105)  & 0.259 (0.098)  & 0.343 (0.098)  & 0.468 (0.067)  \\
 &  SVM & 0.325 (0.138)  & 0.377 (0.063)  & 0.391 (0.101)  & 0.431 (0.080)  & 0.332 (0.114)  & 0.468 (0.083)  \\
 &  LSTM & 0.489 (0.090)  & 0.373 (0.086)  & 0.450 (0.084)  & 0.468 (0.079)  & 0.423 (0.057)  & 0.579 (0.099)  \\
 &  biLSTM & \textbf{0.722} (0.057)  & \textbf{0.481} (0.097)  & \textbf{0.522} (0.080)  & \textbf{0.562} (0.086)  & \textbf{0.551} (0.086)  & \textbf{0.669} (0.064)  \\
\hline
TSS  &  RF & 0.413 (0.090)  & 0.344 (0.094)  & 0.437 (0.072)  & 0.376 (0.101)  & 0.426 (0.061)  & 0.579 (0.085)  \\
 &  MLP & 0.326 (0.123)  & 0.281 (0.125)  & 0.379 (0.098)  & 0.283 (0.101)  & 0.382 (0.092)  & 0.504 (0.046)  \\
 &  SVM & 0.336 (0.140)  & 0.413 (0.045)  & 0.417 (0.091)  & 0.459 (0.063)  & 0.374 (0.106)  & 0.507 (0.057)  \\
 &  LSTM & 0.501 (0.093)  & 0.407 (0.071)  & 0.487 (0.059)  & 0.499 (0.063)  & 0.468 (0.051)  & 0.615 (0.082)  \\
 &  biLSTM & \textbf{0.737} (0.057)  & \textbf{0.515} (0.081)  & \textbf{0.567} (0.059)  & \textbf{0.592} (0.073)  & \textbf{0.590} (0.063)  & \textbf{0.703} (0.041)  \\
\hline
WAUC  &  RF & 0.453 (0.056)  & 0.375 (0.071)  & 0.476 (0.048)  & 0.410 (0.063)  & 0.459 (0.040)  & 0.621 (0.057)  \\
 &  MLP & 0.354 (0.033)  & 0.301 (0.032)  & 0.415 (0.088)  & 0.304 (0.072)  & 0.410 (0.024)  & 0.543 (0.085)  \\
 &  SVM & 0.361 (0.087)  & 0.453 (0.068)  & 0.457 (0.023)  & 0.503 (0.064)  & 0.405 (0.040)  & 0.553 (0.085)  \\
 &  LSTM & 0.541 (0.074)  & 0.436 (0.071)  & 0.526 (0.020)  & 0.535 (0.052)  & 0.510 (0.051)  & 0.671 (0.026)  \\
 &  biLSTM & \textbf{0.794} (0.041)  & \textbf{0.563} (0.039)  & \textbf{0.609} (0.086)  & \textbf{0.646} (0.043)  & \textbf{0.642} (0.048)  & \textbf{0.764} (0.073)  \\
	\enddata
\end{deluxetable*}
\begin{deluxetable*}{llcccccc}[ht!]
	\tablecaption{Performance Comparison of RF, MLP, SVM, LSTM and biLSTM 
	Based on Our Cross-Validation Scheme
	for the F\_S Problem 
\label{tab:seppredictionresultsf}}
\tablewidth{0pt}
	\tablehead{
		\colhead{}&{}& \colhead{12 hr} & \colhead{24 hr} & \colhead{36 hr} &
		\colhead{48 hr} & \colhead{60 hr} &\colhead{72 hr}
	}
	\startdata
Recall  &  RF & 0.414 (0.099)  & 0.468 (0.066)  & 0.592 (0.137)  & 0.550 (0.123)  & 0.522 (0.120)  & 0.590 (0.109)  \\
 &  MLP & 0.367 (0.138)  & 0.457 (0.098)  & 0.398 (0.172)  & 0.501 (0.148)  & 0.520 (0.153)  & 0.568 (0.150)  \\
 &  SVM & 0.433 (0.064)  & 0.471 (0.058)  & 0.495 (0.189)  & 0.575 (0.106)  & 0.518 (0.160)  & 0.567 (0.145)  \\
 &  LSTM & 0.468 (0.146)  & 0.456 (0.166)  & 0.592 (0.155)  &
 \textbf{0.591} (0.140)  & 0.546 (0.155)  & 0.624 (0.127)  \\
 &  biLSTM & \textbf{0.520} (0.103)  & \textbf{0.508} (0.122)  & \textbf{0.597} (0.095)  & 0.545 (0.197)  & \textbf{0.548} (0.090)  & \textbf{0.629} (0.120)  \\
\hline
Precision  &  RF & 0.178 (0.052)  & 0.253 (0.119)  & 0.267 (0.159)  & 0.314 (0.144)  & 0.285 (0.166)  & 0.370 (0.177)  \\
 &  MLP & 0.164 (0.073)  & 0.215 (0.106)  & 0.227 (0.140)  & 0.254 (0.136)  & 0.247 (0.138)  & 0.315 (0.158)  \\
 &  SVM & 0.152 (0.035)  & 0.218 (0.096)  & 0.278 (0.130)  & 0.287 (0.129)  & 0.234 (0.136)  & 0.306 (0.154)  \\
 &  LSTM & 0.184 (0.076)  & 0.252 (0.114)  & 0.432 (0.124)  & 0.340 (0.141)  & 0.329 (0.162)  & 0.404 (0.161)  \\
 &  biLSTM & \textbf{0.366} (0.159)  & \textbf{0.473} (0.110)  & \textbf{0.527} (0.155)  & \textbf{0.377} (0.200)  & \textbf{0.405} (0.170)  & \textbf{0.485} (0.166)  \\
\hline
BACC  &  RF & 0.627 (0.033)  & 0.656 (0.030)  & 0.684 (0.052)  & 0.681 (0.038)  & 0.650 (0.025)  & 0.702 (0.036)  \\
 &  MLP & 0.599 (0.049)  & 0.635 (0.031)  & 0.605 (0.062)  & 0.634 (0.036)  & 0.618 (0.040)  & 0.665 (0.032)  \\
 &  SVM & 0.616 (0.052)  & 0.641 (0.028)  & 0.655 (0.039)  & 0.676 (0.035)  & 0.605 (0.055)  & 0.654 (0.052)  \\
 &  LSTM & 0.647 (0.046)  & 0.653 (0.079)  & 0.739 (0.061)  & 0.703 (0.022)  & 0.677 (0.059)  & 0.720 (0.029)  \\
 &  biLSTM & \textbf{0.721} (0.046)  & \textbf{0.714} (0.047)  & \textbf{0.765} (0.037)  & \textbf{0.706} (0.076)  & \textbf{0.708} (0.015)  & \textbf{0.754} (0.027)  \\
\hline
HSS  &  RF & 0.154 (0.031)  & 0.218 (0.066)  & 0.239 (0.117)  & 0.265 (0.077)  & 0.212 (0.069)  & 0.311 (0.086)  \\
 &  MLP & 0.128 (0.073)  & 0.171 (0.049)  & 0.156 (0.098)  & 0.181 (0.050)  & 0.152 (0.048)  & 0.235 (0.049)  \\
 &  SVM & 0.119 (0.040)  & 0.175 (0.038)  & 0.233 (0.082)  & 0.235 (0.042)  & 0.128 (0.064)  & 0.218 (0.087)  \\
 &  LSTM & 0.170 (0.064)  & 0.216 (0.091)  & 0.414 (0.110)  & 0.303 (0.068)  & 0.270 (0.100)  & 0.352 (0.078)  \\
 &  biLSTM & \textbf{0.365} (0.137)  & \textbf{0.418} (0.100)  & \textbf{0.493} (0.099)  & \textbf{0.345} (0.159)  & \textbf{0.353} (0.079)  & \textbf{0.448} (0.089)  \\
\hline
TSS  &  RF & 0.254 (0.066)  & 0.313 (0.060)  & 0.368 (0.103)  & 0.362 (0.075)  & 0.301 (0.050)  & 0.405 (0.071)  \\
 &  MLP & 0.198 (0.098)  & 0.270 (0.062)  & 0.211 (0.125)  & 0.269 (0.072)  & 0.236 (0.080)  & 0.329 (0.064)  \\
 &  SVM & 0.233 (0.104)  & 0.282 (0.057)  & 0.309 (0.078)  & 0.353 (0.070)  & 0.210 (0.110)  & 0.309 (0.104)  \\
 &  LSTM & 0.293 (0.092)  & 0.305 (0.158)  & 0.479 (0.122)  & 0.405 (0.044)  & 0.353 (0.119)  & 0.440 (0.058)  \\
 &  biLSTM & \textbf{0.441} (0.093)  & \textbf{0.428} (0.093)  & \textbf{0.529} (0.075)  & \textbf{0.412} (0.151)  & \textbf{0.416} (0.031)  & \textbf{0.509} (0.055)  \\
\hline
WAUC  &  RF & 0.276 (0.022)  & 0.338 (0.067)  & 0.403 (0.021)  & 0.388 (0.035)  & 0.330 (0.021)  & 0.431 (0.042)  \\
 &  MLP & 0.212 (0.084)  & 0.290 (0.079)  & 0.226 (0.024)  & 0.294 (0.034)  & 0.256 (0.085)  & 0.359 (0.075)  \\
 &  SVM & 0.254 (0.084)  & 0.307 (0.048)  & 0.334 (0.062)  & 0.381 (0.074)  & 0.230 (0.060)  & 0.334 (0.027)  \\
 &  LSTM & 0.319 (0.056)  & 0.328 (0.047)  & 0.514 (0.033)  & 0.432 (0.038)  & 0.382 (0.073)  & 0.474 (0.020)  \\
 &  biLSTM & \textbf{0.480} (0.076)  & \textbf{0.467} (0.034)  & \textbf{0.574} (0.085)  & \textbf{0.450} (0.035)  & \textbf{0.448} (0.087)  & \textbf{0.552} (0.016)  \\
	\enddata
\end{deluxetable*}
\subsection{Probabilistic Forecasting and Calibration}
\label{sec:probabilisticprediction}
The five machine learning methods 
(RF, MLP, SVM, LSTM, biLSTM) 
studied here are inherently 
probabilistic forecasting models in the sense that
they calculate a probability between 0 and 1.
We compare the probability with 
a threshold, which is set to 0.5, to determine the output produced
 by each machine learning method.
The output is either 1 or 0 (see Figure \ref{fig:biLSTM_arch}),
and hence each method is essentially a binary prediction model.
In addition to comparing the methods used as binary prediction models,
we also compare the methods used as probabilistic forecasting models,
where the output produced by each model is interpreted as follows.
{\bf [FC\_S problem]}
Given a data sample $x_t$ at time point $t$ 
in an AR where the AR will produce an M- or X-class flare 
within the next $T$ hours of $t$ and the flare initiates a CME, 
based on the 
SHARP
parameters in $x_t$ and its preceding $m-1$ data samples
$x_{t-m+1}$, $x_{t-m+2}$, $\ldots$, $x_{t-1}$,
we calculate and output a probabilistic estimate of how
likely it is that the AR will produce an SEP event associated with the flare and CME.
{\bf [F\_S problem]} 
Given a data sample $x_t$ at time point $t$ in an AR 
where the AR will produce an M- or X-class flare within the next $T$ hours of $t$ 
regardless of whether or not the flare initiates a CME, 
based on the 
SHARP
parameters in $x_t$ and its preceding $m-1$ data samples
$x_{t-m+1}$, $x_{t-m+2}$, $\ldots$, $x_{t-1}$,
we calculate and output a probabilistic estimate of how
likely it is that the AR will produce an SEP event associated with the flare.

The distribution and behavior of the predicted probabilistic values
may not match the expected distribution of observed probabilities in the training data.
One can adjust the distribution of the predicted probabilities to better 
match the expected distribution observed in the training data through calibration.
Here, we adopt isotonic regression~ \citep{Isotonic1967,IsotonicSager1982}
to adjust the probabilities.
Isotonic regression works by fitting a free-form line to a sequence of data points 
such that the fitted line is non-decreasing (or non-increasing) everywhere, 
and lies as close to the data points as possible.
Calibrated models often produce more accurate results.
We add a suffix ``+C'' to each model to denote the calibrated version of the model.

To quantitatively assess the performance of a probabilistic forecasting model, we adopt
the Brier Score \citep[BS;][]{BSSWilks2010} and
Brier Skill Score \citep[BSS;][]{BSSWilks2010}, defined as follows: 
\begin{equation}
\text{BS} = \frac{1}{N}\sum_{i=1}^{N}(y_i - \hat{y}_i)^2,
\end{equation}
\begin{equation}
\text{BSS} = 1 - \frac{BS}{\frac{1}{N}\sum_{i=1}^{N}(y_i - \bar{y})^2}.
\end{equation}
Here, $N$ is the total number of data sequences 
each having $m$ consecutive data samples in
the test set (see Figure \ref{fig:biLSTM_arch} 
where a test data sequence with $m$ consecutive data samples
is fed to our biLSTM model);
$y_i$ denotes the observed probability and $\hat{y}_i$
denotes the predicted probability 
of the $i$th test data sequence respectively;
$\bar{y} = \frac{1}{N}\sum_{i=1}^{N} y_i$ denotes the 
mean of all the observed probabilities.
The BS values range
from 0 to 1, with 0 being a perfect score, whereas the BSS values range from
$-\infty$ to 1, with 1 being a perfect score.

Table \ref{tab:bsbssfc} 
compares the performance of the five machine learning methods used
as probabilistic forecasting models
for the FC\_S and F\_S problems respectively. 
The table presents the mean BS and BSS values
averaged over the six runs based on our cross-validation scheme
with standard deviations enclosed in parentheses.
Best BS and BSS values are highlighted in boldface.
It can be seen from Table \ref{tab:bsbssfc} that 
the probabilistic forecasting models generally perform better in solving the
FC\_S problem than in solving the F\_S problem, suggesting that F\_S is a harder problem
and hence the forecasting results for the F\_S problem would be less reliable.
These findings are consistent with those in Tables \ref{tab:seppredictionresultsfc} and \ref{tab:seppredictionresultsf} where the machine learning methods are used as binary prediction models.
Furthermore, the calibrated version of a model is better than the model without calibration.
Overall, biLSTM+C performs the best among all the models
in terms of both BS and BSS.
\begin{deluxetable*}{lllcccccc}[ht!]
\tablecaption{Probabilistic Forecasting Results of RF, MLP, SVM, LSTM and biLSTM With and Without 
 Calibration for the FC\_S and F\_S Problems Respectively
		\label{tab:bsbssfc}}
\tablewidth{0pt}
	\tablehead{
		\colhead{}&{}&{}& \colhead{12 hr} & \colhead{24 hr} & \colhead{36 hr} &
		\colhead{48 hr} & \colhead{60 hr} &\colhead{72 hr}
	}
	\startdata
FC\_S& BS  &  RF & 0.372 (0.083)  & 0.342 (0.075)  & 0.365 (0.060)  & 0.342 (0.092)  & 0.355 (0.051)  & 0.269 (0.040)  \\ 
&&  RF+C & 0.332 (0.070)  & 0.331 (0.101)  & 0.324 (0.056)  & 0.302 (0.085)  & 0.328 (0.047)  & 0.252 (0.037)  \\
 &&  MLP & 0.362 (0.136)  & 0.393 (0.080)  & 0.335 (0.095)  & 0.315 (0.050)  & 0.335 (0.083)  & 0.280 (0.025)  \\ 
&&  MLP+C & 0.329 (0.124)  & 0.366 (0.076)  & 0.309 (0.088)  & 0.283 (0.050)  & 0.301 (0.073)  & 0.255 (0.027)  \\
 &&  SVM & 0.359 (0.052)  & 0.344 (0.037)  & 0.337 (0.075)  & 0.353 (0.049)  & 0.306 (0.087)  & 0.298 (0.034)  \\ 
&&  SVM+C & 0.322 (0.047)  & 0.303 (0.030)  & 0.297 (0.066)  & 0.306 (0.035)  & 0.284 (0.080)  & 0.267 (0.035)  \\
 &&  LSTM & 0.337 (0.064)  & 0.356 (0.054)  & 0.300 (0.058)  & 0.288 (0.032)  & 0.298 (0.038)  & 0.274 (0.028)  \\ 
&&  LSTM+C & 0.271 (0.062)  & 0.302 (0.045)  & 0.262 (0.053)  & 0.244 (0.032)  & 0.273 (0.035)  & 0.232 (0.021)  \\
 &&  biLSTM & 0.248 (0.028)  & 0.272 (0.052)  & 0.270 (0.048)  & 0.281 (0.040)  & 0.297 (0.021)  & 0.279 (0.015)  \\ 
& & biLSTM+C & \textbf{0.215} (0.021)  & \textbf{0.249} (0.038)  & \textbf{0.223} (0.025)  & \textbf{0.235} (0.033)  & \textbf{0.270} (0.043)  & \textbf{0.202} (0.014)  \\
 \cline{2-9}
&BSS  &  RF & 0.273 (0.166)  & 0.316 (0.164)  & 0.282 (0.124)  & 0.316 (0.180)  & 0.325 (0.118)  & 0.466 (0.067)  \\ 
&&  RF+C & 0.341 (0.140)  & 0.343 (0.189)  & 0.362 (0.115)  & 0.396 (0.167)  & 0.340 (0.109)  & 0.501 (0.062)  \\
 &&  MLP & 0.290 (0.262)  & 0.274 (0.100)  & 0.320 (0.202)  & 0.382 (0.082)  & 0.323 (0.179)  & 0.436 (0.048)  \\ 
& & MLP+C & 0.355 (0.239)  & 0.325 (0.094)  & 0.372 (0.187)  & 0.445 (0.087)  & 0.392 (0.158)  & 0.486 (0.052)  \\
 & & SVM & 0.281 (0.128)  & 0.310 (0.086)  & 0.333 (0.142)  & 0.295 (0.101)  & 0.388 (0.178)  & 0.406 (0.057)  \\ 
&&  SVM+C & 0.355 (0.115)  & 0.392 (0.065)  & 0.412 (0.125)  & 0.389 (0.074)  & 0.432 (0.164)  & 0.469 (0.059)  \\
 &&  LSTM & 0.338 (0.124)  & 0.306 (0.114)  & 0.388 (0.128)  & 0.425 (0.073)  & 0.406 (0.074)  & 0.458 (0.052)  \\ 
&&  LSTM+C & 0.466 (0.121)  & 0.395 (0.099)  & 0.466 (0.115)  & 0.513 (0.068)  & 0.456 (0.068)  & 0.542 (0.037)  \\
 &&  biLSTM & 0.513 (0.050)  & 0.450 (0.110)  & 0.464 (0.097)  & 0.424 (0.086)  & 0.417 (0.046)  & 0.450 (0.018)  \\ 
& & biLSTM+C & \textbf{0.578} (0.035)  & \textbf{0.498} (0.080)  & \textbf{0.558} (0.046)  & \textbf{0.518} (0.065)  & \textbf{0.470} (0.087)  & \textbf{0.587} (0.020)  \\
\hline
F\_S&
BS  &  RF & 0.393 (0.094)  & 0.391 (0.075)  & 0.449 (0.126)  & 0.459 (0.077)  & 0.381 (0.063)  & 0.317 (0.056)  \\ 
& &  RF+C & 0.341 (0.078)  & 0.351 (0.062)  & 0.380 (0.109)  & 0.383 (0.063)  & 0.334 (0.043)  & 0.276 (0.044)  \\
  &&  MLP & 0.433 (0.042)  & 0.429 (0.053)  & 0.395 (0.063)  & 0.404 (0.105)  & 0.394 (0.134)  & 0.366 (0.072)  \\ 
& &  MLP+C & 0.376 (0.031)  & 0.381 (0.046)  & 0.340 (0.057)  & 0.357 (0.100)  & 0.341 (0.111)  & 0.329 (0.069)  \\
  &&  SVM & 0.429 (0.071)  & 0.391 (0.043)  & 0.381 (0.032)  & 0.379 (0.075)  & 0.403 (0.067)  & 0.390 (0.100)  \\ 
& &  SVM+C & 0.390 (0.073)  & 0.354 (0.038)  & 0.336 (0.025)  & 0.336 (0.061)  & 0.363 (0.067)  & 0.346 (0.082)  \\
 & &  LSTM & 0.373 (0.093)  & 0.359 (0.077)  & 0.381 (0.048)  & 0.347 (0.042)  & 0.377 (0.074)  & 0.276 (0.034)  \\ 
& &  LSTM+C & 0.341 (0.088)  & 0.315 (0.074)  & 0.336 (0.049)  & 0.314 (0.036)  & 0.319 (0.052)  & 0.247 (0.034)  \\
  &&  biLSTM & 0.267 (0.054)  & 0.345 (0.038)  & 0.318 (0.069)  & 0.344 (0.040)  & 0.346 (0.034)  & 0.307 (0.028)  \\ 
 &&  biLSTM+C & \textbf{0.231} (0.042)  & \textbf{0.294} (0.035)  & \textbf{0.226} (0.060)  & \textbf{0.291} (0.044)  & \textbf{0.289} (0.021)  & \textbf{0.220} (0.029)  \\
\cline{2-9}
&
BSS  &  RF & 0.267 (0.182)  & 0.206 (0.160)  & 0.232 (0.224)  & 0.228 (0.066)  & 0.313 (0.131)  & 0.360 (0.093)  \\ 
 & &  RF+C & 0.322 (0.150)  & 0.329 (0.133)  & 0.293 (0.216)  & 0.354 (0.078)  & 0.322 (0.093)  & 0.441 (0.072)  \\
 &  &  MLP & 0.282 (0.078)  & 0.138 (0.109)  & 0.226 (0.105)  & 0.284 (0.076)  & 0.259 (0.222)  & 0.345 (0.085)  \\ 
 & &  MLP+C & 0.336 (0.059)  & 0.235 (0.093)  & 0.335 (0.088)  & 0.368 (0.069)  & 0.358 (0.183)  & 0.411 (0.086)  \\
  & &  SVM & 0.122 (0.163)  & 0.228 (0.083)  & 0.251 (0.051)  & 0.247 (0.151)  & 0.204 (0.137)  & 0.310 (0.119)  \\ 
& &   SVM+C & 0.201 (0.165)  & 0.301 (0.075)  & 0.339 (0.043)  & 0.332 (0.122)  & 0.283 (0.135)  & 0.388 (0.091)  \\
  & &  LSTM & 0.256 (0.204)  & 0.297 (0.150)  & 0.230 (0.092)  & 0.309 (0.080)  & 0.342 (0.152)  & 0.430 (0.129)  \\ 
& &   LSTM+C & 0.319 (0.193)  & 0.383 (0.144)  & 0.323 (0.096)  & 0.385 (0.067)  & 0.447 (0.145)  & 0.489 (0.122)  \\
 & &   biLSTM & 0.464 (0.106)  & 0.309 (0.092)  & 0.451 (0.144)  & 0.314 (0.083)  & 0.351 (0.072)  & 0.437 (0.091)  \\ 
& &   biLSTM+C & \textbf{0.535} (0.083)  & \textbf{0.410} (0.084)  & \textbf{0.513} (0.100)  & \textbf{0.420} (0.087)  & \textbf{0.457} (0.047)  & \textbf{0.521} (0.089)  \\
\enddata
\end{deluxetable*}

\section{Discussion and Conclusions} 
\label{sec:conclusions}
We develop a bidirectional long short-term memory (biLSTM) network for SEP prediction.
We consider two prediction tasks.
In the first task (FC\_S),
given a data sample $x_t$ at time point $t$ 
in an AR where the AR will produce an M- or X-class flare within the next $T$ hours of $t$ and the flare initiates a CME, based on the 
SHARP
parameters in $x_t$ and its preceding $m-1$ data samples
$x_{t-m+1}$, $x_{t-m+2}$, $\ldots$, $x_{t-1}$,
our biLSTM, when used as a binary prediction
model, can predict whether the AR will produce an SEP event 
associated with the flare/CME.
Furthermore, our biLSTM, when used as
a probabilistic forecasting model, can provide a probabilistic
estimate of how likely it is that the AR will produce an SEP event associated with
the flare/CME.
In the second task (F\_S), 
given a data sample $x_t$ at time point $t$ in an AR 
where the AR will produce an M- or X-class flare within the next $T$ hours of $t$,
based on the 
SHARP
parameters in $x_t$ and its preceding $m-1$ data samples
$x_{t-m+1}$, $x_{t-m+2}$, $\ldots$, $x_{t-1}$,
our biLSTM, when used as a binary prediction
model, can predict whether the AR will produce an SEP event 
associated with the flare, and when used as 
a probabilistic forecasting model,
can provide a probabilistic
estimate of how likely it is that the AR will produce an SEP event associated with
the flare, 
regardless of whether or not the flare initiates a CME.
For both tasks, $T$ ranges from 12 to 72 in
12 hr intervals.

We surveyed and collected data samples from the JSOC website, in the period between 2010 and 2021.
Each data sample contains 18 SHARP parameters.
Active regions (ARs) from 2010, 2016, and 2018-2021 were excluded from the study 
due to the lack of qualified data samples or
the absence of SEP events associated with M-/X-class flares and CMEs.
We then performed a cross-validation study on the remaining six years (2011-2015 and 2017).
In the cross-validation study,
training and test sets are disjoint, and hence 
our biLSTM model can make predictions on ARs that were never seen before. 
We evaluated the performance of our model
and compared it with four related machine learning algorithms,
namely RF \citep{Liu..Wang..Solar..2017ApJ...843..104L}, MLP \citep{Inceoglu_2018CMESEP}, SVM \citep{BobraCME2016ApJ...821..127B} and a previous LSTM network \citep{Liu_2019FlarePrediction}.
The five machine learning methods including our biLSTM
can be used both as binary prediction models and as probabilistic forecasting models.
Our main results are summarized
as follows.
\begin{enumerate}
	\item {
	The data samples in an AR are modeled as time series.
	We employ the biLSTM network to predict SEP events based on the time series.
	To our knowledge, this is the first study using a deep neural network to 
	learn the dependencies in the temporal domain of the data
	for SEP prediction.
	}
	\item {
	We evaluate the importance of the 18 SHARP parameters used in our study. 
	It is found that using the top 15 SHARP parameters achieves the best performance 
	for both the FC\_S and F\_S tasks.
	This finding is consistent with the literature which indicates using fewer high-quality SHARP parameters often achieves better performance for eruption prediction than using all the SHARP parameters including low-quality ones
	\citep{MLNewAIAlpaydin2016,BobraCME2016ApJ...821..127B,Liu_2020CMEPrediction}.
	}
	\item {
	Our experiments show that the proposed biLSTM outperforms the four related 
	machine learning methods in performing binary prediction and probabilistic forecasting
	for both the FC\_S and F\_S tasks.
	Furthermore, we introduce a calibration mechanism to enhance the accuracy of probabilistic forecasting. 
	Overall, the calibrated biLSTM achieves the best performance among all the probabilistic forecasting models studied here.
	}
	\item {
	When both an M-/X-class flare and its associated CME will occur, predicting whether there is an SEP event
	associated with the flare and CME is an easier problem (FC\_S).
	Our biLSTM can solve the FC\_S problem with relatively high accuracy.
	In contrast, when an M-/X-class flare will occur in the absence of CME information, 
	predicting whether there is an SEP event associated with the flare is a harder problem (F\_S).
	Our biLSTM solves the F\_S problem with relatively low accuracy, 
	and hence the prediction results would be less reliable.
	}
	\item {
 The findings reported here are based on
    the cross-validation (CV) scheme in which 
     six years (2011-2015 and 2017) are considered,
    data samples from each year in turn are used for testing, and
    data samples from the other five years together are used for training.
     To further understand the behavior of our biLSTM network and the four related machine learning methods, 
     we have performed additional
     experiments using a random division (RD) scheme.
     With RD, we randomly select 10\% of all positive data sequences and 
      10\% of all negative data sequences, and use them together as the test set. 
     The remaining 90\% of the positive data sequences and 90\% of the negative data sequences 
     are used together as the training set. We repeat this experiment 100 times.
     The average values and standard deviations of the performance metrics are calculated.
     Tables \ref{tab:seppredictionresultsfcrd} and \ref{tab:seppredictionresultsfrd} in the Appendix present
     results of the five machine learning methods used as binary prediction models for the FC\_S and F\_S problems respectively.
     Table \ref{tab:bsbssrd} presents results of the five machine learning methods used as probabilistic forecasting models for the FC\_S and F\_S problems respectively.
     It can be seen from these tables that the results obtained from the random division scheme
     are consistent with those from the cross-validation scheme,
     though the performance metric values from the RD scheme are generally better than those from the CV scheme.
     This happens probably because with the RD scheme the machine learning methods are trained by more diverse data and hence are more knowledgeable, yielding more accurate results than with the CV scheme.
	}
\end{enumerate}

It should be pointed out that,
in solving the FC\_S problem,
the condition in which 
we have a data sample $x_t$ at time point $t$ 
in an AR where the AR will produce an M- or X-class flare 
within the next $T$ hours of $t$ and the flare initiates a CME
is given.
That is,
we assume an M- or X-class flare and its associated CME will occur.
In an operational system, one can determine in two phases if an AR will produce an M- or X-class flare within the next $T$-hours of a given time point $t$ and
if the flare initiates a CME, as follows \citep{HaoDisseration}.
In the first phase, one can use a flare prediction tool 
\citep[e.g.,][]{Liu..Wang..Solar..2017ApJ...843..104L,ForecastingFlaresPredictors2018SoPh..293...28F,2018SoPh..293...48J,2018ApJ...858..113N_Nishizuka,Liu_2019FlarePrediction}
to predict whether there will be an
M- or X-class flare within the next $T$ hours of $t$. 
If the answer is yes, then in the second 
phase one can use a CME prediction tool
\citep[e.g.,][]{Liu_2020CMEPrediction} to predict 
whether the flare initiates a CME.
If the answer is also yes, then one can use the proposed biLSTM
to predict whether there is an SEP event associated with the flare and CME.
On the other hand, to solve the F\_S problem,
one only needs to execute the 
first phase.
If the answer from the first phase
indicates that an M- or X-class flare will occur within the next $T$ hours of $t$,
one can then go ahead to use the proposed biLSTM to predict whether there is an SEP event associated 
with the flare.
Thus, 
the proposed biLSTM does not
function in a stand-alone manner.
Rather, it first requires the other tools to provide flare/CME predictions.
As such, the performance of the operational biLSTM system
depends on the performance of the other tools.
A wrong prediction from the other tools would
affect the accuracy of our approach.

We thank the referee and scientific editor for very helpful
and thoughtful comments.
We also thank the team of \textit{SDO}/HMI for producing 
vector magnetic data products.
The flare catalogs were prepared by and made available through NOAA NCEI.
The CME and SEP event records were provided by DONKI.
This work was supported by U.S. NSF grants AGS-1927578 and AGS-1954737.
J.W. thanks Manolis K. Georgoulis for helpful conversations in the SHINE 2019 Conference.
Q.L. and H.W. acknowledge the support of NASA under grants 
80NSSC18K1705,
80NSSC19K0068
and
80NSSC20K1282.

\appendix
\vspace{-0.4cm}
Tables \ref{tab:seppredictionresultsfcrd} and \ref{tab:seppredictionresultsfrd} present results
     of the five machine learning methods 
     (RF, MLP, SVM, LSTM, biLSTM) 
     used as binary prediction models for the FC\_S and F\_S problems respectively.
     Table \ref{tab:bsbssrd} presents results of the five machine learning methods used as probabilistic forecasting models for the FC\_S and F\_S problems respectively.
     The tables show the mean performance metric values averaged over the 100 experiments based on 
    the random division scheme with standard deviations enclosed in parentheses.
    Best average metric values are highlighted in boldface.
    
\begin{deluxetable*}{llcccccc}[ht!]
	\tablecaption{Performance Comparison of RF, MLP, SVM, LSTM and biLSTM Based on the Random Division Scheme for the FC\_S Problem}
	 \label{tab:seppredictionresultsfcrd}
\tablewidth{0pt}
	\tablehead{
		\colhead{}&{}& \colhead{12 hr} & \colhead{24 hr} & \colhead{36 hr} &
		\colhead{48 hr} & \colhead{60 hr} &\colhead{72 hr}
	}
	\startdata
Recall  &  RF & 0.699 (0.149)  & 0.686 (0.148)  & 0.657 (0.169)  & 0.701 (0.165)  & 0.689 (0.087)  & 0.817 (0.080)  \\
 &  MLP & 0.655 (0.164)  & 0.690 (0.132)  & 0.688 (0.149)  & 0.692 (0.149)  & 0.679 (0.101)  & 0.798 (0.091)  \\
 &  SVM & 0.666 (0.170)  & 0.693 (0.147)  & 0.681 (0.159)  & 0.701 (0.153)  & 0.716 (0.088)  & 0.810 (0.091)  \\
 &  LSTM & 0.786 (0.080)  & 0.804 (0.071)  & 0.840 (0.059)  & 0.860 (0.047)  & 0.870 (0.054)  & 0.884 (0.061)  \\
 &  biLSTM & \textbf{0.911} (0.056)  & \textbf{0.844} (0.067)  & \textbf{0.860} (0.057)  & \textbf{0.882} (0.047)  & \textbf{0.875} (0.052)  & \textbf{0.906} (0.054)  \\
\hline
Precision  &  RF & 0.680 (0.184)  & 0.670 (0.113)  & 0.654 (0.069)  & 0.636 (0.088)  & 0.675 (0.036)  & 0.719 (0.046)  \\
 &  MLP & 0.650 (0.214)  & 0.630 (0.079)  & 0.659 (0.069)  & 0.627 (0.073)  & 0.659 (0.040)  & 0.715 (0.045)  \\
 &  SVM & 0.733 (0.163)  & 0.649 (0.090)  & 0.707 (0.083)  & 0.684 (0.093)  & 0.692 (0.042)  & 0.725 (0.046)  \\
 &  LSTM & 0.682 (0.135)  & 0.692 (0.130)  & 0.706 (0.103)  & 0.683 (0.120)  & 0.710 (0.106)  & 0.727 (0.071)  \\
 &  biLSTM & \textbf{0.788} (0.149)  & \textbf{0.721} (0.132)  & \textbf{0.722} (0.103)  & \textbf{0.705} (0.118)  & \textbf{0.752} (0.107)  & \textbf{0.749} (0.067)  \\
\hline
BACC  &  RF & 0.781 (0.073)  & 0.776 (0.068)  & 0.761 (0.078)  & 0.768 (0.083)  & 0.770 (0.042)  & 0.831 (0.042)  \\
 &  MLP & 0.750 (0.085)  & 0.768 (0.057)  & 0.774 (0.069)  & 0.759 (0.067)  & 0.761 (0.049)  & 0.822 (0.046)  \\
 &  SVM & 0.787 (0.093)  & 0.775 (0.065)  & 0.787 (0.080)  & 0.784 (0.080)  & 0.787 (0.046)  & 0.831 (0.047)  \\
 &  LSTM & 0.822 (0.051)  & 0.828 (0.051)  & 0.847 (0.044)  & 0.840 (0.054)  & 0.848 (0.042)  & 0.860 (0.046)  \\
 &  biLSTM & \textbf{0.906} (0.041)  & \textbf{0.854} (0.048)  & \textbf{0.861} (0.042)  & \textbf{0.858} (0.051)  & \textbf{0.867} (0.045)  & \textbf{0.878} (0.040)  \\
\hline
HSS  &  RF & 0.548 (0.152)  & 0.541 (0.124)  & 0.516 (0.130)  & 0.516 (0.150)  & 0.536 (0.072)  & 0.639 (0.076)  \\
 &  MLP & 0.492 (0.183)  & 0.517 (0.096)  & 0.537 (0.118)  & 0.500 (0.119)  & 0.516 (0.083)  & 0.624 (0.081)  \\
 &  SVM & 0.585 (0.178)  & 0.534 (0.110)  & 0.575 (0.141)  & 0.561 (0.148)  & 0.568 (0.080)  & 0.640 (0.082)  \\
 &  LSTM & 0.612 (0.128)  & 0.624 (0.125)  & 0.656 (0.105)  & 0.633 (0.131)  & 0.656 (0.108)  & 0.682 (0.093)  \\
 &  biLSTM & \textbf{0.769} (0.124)  & \textbf{0.670} (0.123)  & \textbf{0.682} (0.103)  & \textbf{0.667} (0.126)  & \textbf{0.702} (0.112)  & \textbf{0.717} (0.083)  \\
 \hline
TSS  &  RF & 0.562 (0.146)  & 0.551 (0.136)  & 0.523 (0.155)  & 0.536 (0.165)  & 0.541 (0.084)  & 0.662 (0.084)  \\
 &  MLP & 0.501 (0.171)  & 0.536 (0.114)  & 0.549 (0.139)  & 0.519 (0.134)  & 0.522 (0.097)  & 0.644 (0.093)  \\
 &  SVM & 0.574 (0.186)  & 0.551 (0.131)  & 0.573 (0.161)  & 0.568 (0.161)  & 0.574 (0.091)  & 0.661 (0.093)  \\
 &  LSTM & 0.645 (0.103)  & 0.657 (0.102)  & 0.695 (0.088)  & 0.680 (0.109)  & 0.697 (0.085)  & 0.720 (0.091)  \\
 &  biLSTM & \textbf{0.812} (0.081)  & \textbf{0.708} (0.096)  & \textbf{0.722} (0.073)  & \textbf{0.715} (0.103)  & \textbf{0.733} (0.091)  & \textbf{0.756} (0.079)  \\
\hline
WAUC  &  RF & 0.619 (0.022)  & 0.601 (0.046)  & 0.577 (0.022)  & 0.579 (0.050)  & 0.599 (0.065)  & 0.729 (0.072)  \\
 &  MLP & 0.551 (0.015)  & 0.581 (0.013)  & 0.605 (0.042)  & 0.573 (0.051)  & 0.565 (0.024)  & 0.709 (0.035)  \\
 &  SVM & 0.633 (0.015)  & 0.597 (0.079)  & 0.630 (0.048)  & 0.618 (0.057)  & 0.625 (0.056)  & 0.730 (0.046)  \\
 &  LSTM & 0.708 (0.039)  & 0.725 (0.020)  & 0.757 (0.084)  & 0.744 (0.081)  & 0.761 (0.029)  & 0.782 (0.070)  \\
 &  biLSTM & \textbf{0.895} (0.013)  & \textbf{0.775} (0.041)  & \textbf{0.799} (0.051)  & \textbf{0.780} (0.069)  & \textbf{0.796} (0.058)  & \textbf{0.821} (0.063)  \\
	\enddata
\end{deluxetable*}

\begin{deluxetable*}{llcccccc}[ht!]
	\tablecaption{Performance Comparison of RF, MLP, SVM, LSTM and biLSTM Based on the Random Division Scheme for the F\_S Problem}
	 \label{tab:seppredictionresultsfrd}
\tablewidth{0pt}
	\tablehead{
		\colhead{}&{}& \colhead{12 hr} & \colhead{24 hr} & \colhead{36 hr} &
		\colhead{48 hr} & \colhead{60 hr} &\colhead{72 hr}
	}
	\startdata
Recall  &  RF & 0.734 (0.105)  & 0.797 (0.117)  & 0.749 (0.147)  & 0.710 (0.116)  & 0.717 (0.089)  & 0.756 (0.104)  \\
 &  MLP & 0.683 (0.087)  & 0.769 (0.123)  & 0.752 (0.109)  & 0.689 (0.121)  & 0.702 (0.103)  & 0.728 (0.083)  \\
 &  SVM & 0.767 (0.107)  & 0.791 (0.101)  & 0.786 (0.104)  & 0.730 (0.120)  & 0.740 (0.103)  & 0.743 (0.086)  \\
 &  LSTM & 0.774 (0.098)  & 0.770 (0.125)  & 0.817 (0.108)  & 0.782 (0.120)  & 0.768 (0.106)  & 0.772 (0.103)  \\
 &  biLSTM & \textbf{0.834} (0.096)  & \textbf{0.837} (0.120)  & \textbf{0.856} (0.107)  & \textbf{0.837} (0.122)  & \textbf{0.818} (0.101)  & \textbf{0.815} (0.102)  \\
\hline
Precision  &  RF & 0.243 (0.117)  & 0.244 (0.103)  & 0.261 (0.096)  & 0.308 (0.089)  & 0.396 (0.190)  & 0.434 (0.162)  \\
 &  MLP & 0.225 (0.107)  & 0.243 (0.100)  & 0.269 (0.082)  & 0.291 (0.069)  & 0.363 (0.156)  & 0.366 (0.097)  \\
 &  SVM & 0.235 (0.104)  & 0.231 (0.086)  & 0.242 (0.067)  & 0.326 (0.099)  & 0.383 (0.166)  & 0.393 (0.113)  \\
 &  LSTM & 0.250 (0.114)  & 0.252 (0.097)  & 0.291 (0.092)  & 0.349 (0.108)  & 0.413 (0.189)  & 0.443 (0.165)  \\
 &  biLSTM & \textbf{0.279} (0.131)  & \textbf{0.275} (0.107)  & \textbf{0.306} (0.099)  & \textbf{0.377} (0.119)  & \textbf{0.483} (0.174)  & \textbf{0.476} (0.173)  \\
\hline
BACC  &  RF & 0.770 (0.073)  & 0.774 (0.058)  & 0.758 (0.087)  & 0.760 (0.054)  & 0.760 (0.083)  & 0.795 (0.049)  \\
 &  MLP & 0.742 (0.065)  & 0.764 (0.058)  & 0.766 (0.054)  & 0.746 (0.056)  & 0.748 (0.069)  & 0.770 (0.045)  \\
 &  SVM & 0.783 (0.069)  & 0.770 (0.057)  & 0.764 (0.071)  & 0.772 (0.056)  & 0.770 (0.069)  & 0.783 (0.046)  \\
 &  LSTM & 0.791 (0.068)  & 0.772 (0.062)  & 0.799 (0.053)  & 0.801 (0.056)  & 0.787 (0.069)  & 0.804 (0.049)  \\
 &  biLSTM & \textbf{0.825} (0.068)  & \textbf{0.809} (0.059)  & \textbf{0.821} (0.053)  & \textbf{0.832} (0.057)  & \textbf{0.841} (0.059)  & \textbf{0.830} (0.049)  \\
\hline
HSS  &  RF & 0.284 (0.151)  & 0.274 (0.129)  & 0.284 (0.128)  & 0.327 (0.102)  & 0.390 (0.196)  & 0.442 (0.153)  \\
 &  MLP & 0.255 (0.138)  & 0.269 (0.125)  & 0.295 (0.098)  & 0.306 (0.089)  & 0.358 (0.166)  & 0.379 (0.110)  \\
 &  SVM & 0.280 (0.138)  & 0.260 (0.111)  & 0.267 (0.101)  & 0.350 (0.110)  & 0.387 (0.174)  & 0.409 (0.122)  \\
 &  LSTM & 0.298 (0.149)  & 0.284 (0.122)  & 0.330 (0.110)  & 0.385 (0.115)  & 0.418 (0.189)  & 0.455 (0.155)  \\
 &  biLSTM & \textbf{0.340} (0.165)  & \textbf{0.321} (0.131)  & \textbf{0.354} (0.116)  & \textbf{0.426} (0.124)  & \textbf{0.515} (0.158)  & \textbf{0.500} (0.162)  \\
\hline
TSS  &  RF & 0.540 (0.145)  & 0.548 (0.116)  & 0.516 (0.174)  & 0.519 (0.109)  & 0.521 (0.166)  & 0.589 (0.098)  \\
 &  MLP & 0.484 (0.130)  & 0.527 (0.116)  & 0.533 (0.108)  & 0.493 (0.111)  & 0.497 (0.138)  & 0.540 (0.090)  \\
 &  SVM & 0.566 (0.138)  & 0.540 (0.113)  & 0.528 (0.142)  & 0.545 (0.111)  & 0.539 (0.138)  & 0.567 (0.093)  \\
 &  LSTM & 0.581 (0.137)  & 0.544 (0.125)  & 0.598 (0.106)  & 0.601 (0.112)  & 0.574 (0.138)  & 0.607 (0.097)  \\
 &  biLSTM & \textbf{0.651} (0.136)  & \textbf{0.617} (0.119)  & \textbf{0.641} (0.106)  & \textbf{0.664} (0.114)  & \textbf{0.682} (0.118)  & \textbf{0.660} (0.097)  \\
\hline
WAUC  &  RF & 0.591 (0.057)  & 0.600 (0.021)  & 0.561 (0.016)  & 0.565 (0.038)  & 0.577 (0.043)  & 0.655 (0.057)  \\
 &  MLP & 0.528 (0.068)  & 0.577 (0.043)  & 0.578 (0.069)  & 0.543 (0.065)  & 0.547 (0.054)  & 0.597 (0.085)  \\
 &  SVM & 0.624 (0.025)  & 0.586 (0.027)  & 0.580 (0.032)  & 0.602 (0.039)  & 0.590 (0.069)  & 0.617 (0.065)  \\
 &  LSTM & 0.634 (0.052)  & 0.595 (0.048)  & 0.665 (0.026)  & 0.666 (0.027)  & 0.625 (0.028)  & 0.673 (0.062)  \\
 &  biLSTM & \textbf{0.712} (0.068)  & \textbf{0.680} (0.082)  & \textbf{0.710} (0.057)  & \textbf{0.730} (0.053)  & \textbf{0.753} (0.031)  & \textbf{0.732} (0.060)  \\
	\enddata
\end{deluxetable*}

\begin{deluxetable*}{lllcccccc}[ht!]
\tablecaption{Probabilistic Forecasting Results of RF, MLP, SVM, LSTM and biLSTM With and Without Calibration Based on the Random Division Scheme for the FC\_S and F\_S Problems Respectively
		\label{tab:bsbssrd}}
\tablewidth{0pt}
	\tablehead{
		\colhead{}&{}&{}& \colhead{12 hr} & \colhead{24 hr} & \colhead{36 hr} &
		\colhead{48 hr} & \colhead{60 hr} &\colhead{72 hr}
	}
	\startdata
FC\_S& 
BS  &  RF & 0.268 (0.068)  & 0.279 (0.076)  & 0.305 (0.102)  & 0.300 (0.119)  & 0.232 (0.043)  & 0.305 (0.040)  \\ 
 & &  RF+C & 0.226 (0.057)  & 0.246 (0.066)  & 0.260 (0.087)  & 0.256 (0.102)  & 0.272 (0.038)  & 0.269 (0.036)  \\
 &  &  MLP & 0.311 (0.101)  & 0.303 (0.084)  & 0.339 (0.118)  & 0.312 (0.119)  & 0.303 (0.046)  & 0.311 (0.066)  \\ 
 & &  MLP+C & 0.265 (0.087)  & 0.259 (0.071)  & 0.288 (0.101)  & 0.266 (0.102)  & 0.283 (0.040)  & 0.293 (0.058)  \\
 &  &  SVM & 0.261 (0.073)  & 0.264 (0.089)  & 0.271 (0.121)  & 0.261 (0.127)  & 0.285 (0.065)  & 0.282 (0.043)  \\ 
 & &  SVM+C & 0.219 (0.061)  & 0.241 (0.076)  & 0.252 (0.103)  & 0.253 (0.109)  & 0.249 (0.058)  & 0.231 (0.040)  \\
 &  &  LSTM & 0.258 (0.041)  & 0.263 (0.041)  & 0.278 (0.035)  & 0.272 (0.044)  & 0.257 (0.034)  & 0.268 (0.036)  \\ 
 & &  LSTM+C & 0.220 (0.037)  & 0.224 (0.037)  & 0.236 (0.031)  & 0.232 (0.040)  & 0.238 (0.031)  & 0.235 (0.033)  \\
 &  &  biLSTM & 0.214 (0.021)  & 0.230 (0.034)  & 0.203 (0.037)  & 0.214 (0.040)  & 0.193 (0.024)  & 0.185 (0.012)  \\ 
 & &  biLSTM+C & \textbf{0.183} (0.019)  & \textbf{0.195} (0.027)  & \textbf{0.153} (0.028)  & \textbf{0.182} (0.035)  & \textbf{0.164} (0.020)  & \textbf{0.157} (0.011)  \\
 \cline{2-9}
&BSS  &  RF & 0.464 (0.138)  & 0.429 (0.154)  & 0.402 (0.207)  & 0.408 (0.235)  & 0.354 (0.099)  & 0.368 (0.092)  \\ 
 & &  RF+C & 0.549 (0.118)  & 0.516 (0.132)  & 0.490 (0.176)  & 0.495 (0.201)  & 0.453 (0.088)  & 0.460 (0.084)  \\
 &  &  MLP & 0.381 (0.205)  & 0.399 (0.169)  & 0.320 (0.243)  & 0.378 (0.241)  & 0.338 (0.096)  & 0.221 (0.139)  \\ 
 & &  MLP+C & 0.474 (0.175)  & 0.488 (0.143)  & 0.422 (0.209)  & 0.469 (0.207)  & 0.439 (0.084)  & 0.335 (0.122)  \\
 &  &  SVM & 0.476 (0.154)  & 0.431 (0.176)  & 0.433 (0.248)  & 0.386 (0.258)  & 0.415 (0.133)  & 0.409 (0.096)  \\ 
 & &  SVM+C & 0.560 (0.129)  & 0.517 (0.151)  & 0.509 (0.210)  & 0.488 (0.222)  & 0.505 (0.120)  & 0.517 (0.087)  \\
 &  &  LSTM & 0.492 (0.085)  & 0.478 (0.082)  & 0.445 (0.074)  & 0.455 (0.092)  & 0.452 (0.077)  & 0.430 (0.079)  \\ 
 & &  LSTM+C & 0.566 (0.075)  & 0.556 (0.072)  & 0.529 (0.067)  & 0.535 (0.086)  & 0.533 (0.067)  & 0.516 (0.070)  \\
 &  &  biLSTM & 0.573 (0.053)  & 0.545 (0.070)  & 0.596 (0.078)  & 0.569 (0.087)  & 0.610 (0.056)  & 0.627 (0.036)  \\ 
 & &  biLSTM+C & \textbf{0.635} (0.047)  & \textbf{0.614} (0.060)  & \textbf{0.696} (0.058)  & \textbf{0.633} (0.077)  & \textbf{0.668} (0.047)  & \textbf{0.683} (0.032)  \\
\hline
F\_S&
BS  &  RF & 0.284 (0.076)  & 0.288 (0.061)  & 0.372 (0.067)  & 0.320 (0.130)  & 0.307 (0.094)  & 0.310 (0.052)  \\ 
& &   RF+C & 0.272 (0.099)  & 0.276 (0.087)  & 0.312 (0.056)  & 0.268 (0.109)  & 0.274 (0.121)  & 0.281 (0.101)  \\
 & &   MLP & 0.332 (0.114)  & 0.329 (0.072)  & 0.394 (0.080)  & 0.352 (0.080)  & 0.316 (0.145)  & 0.332 (0.147)  \\ 
& &   MLP+C & 0.283 (0.098)  & 0.278 (0.063)  & 0.336 (0.071)  & 0.279 (0.123)  & 0.299 (0.124)  & 0.305 (0.124)  \\
  & &  SVM & 0.276 (0.067)  & 0.279 (0.063)  & 0.298 (0.059)  & 0.276 (0.054)  & 0.295 (0.058)  & 0.290 (0.045)  \\ 
& &   SVM+C & 0.236 (0.059)  & 0.249 (0.055)  & 0.264 (0.050)  & 0.267 (0.045)  & 0.291 (0.049)  & 0.248 (0.038)  \\
  & &  LSTM & 0.277 (0.065)  & 0.279 (0.059)  & 0.285 (0.051)  & 0.282 (0.080)  & 0.273 (0.066)  & 0.289 (0.046)  \\ 
& &   LSTM+C & 0.236 (0.056)  & 0.231 (0.051)  & 0.242 (0.044)  & 0.246 (0.068)  & 0.247 (0.058)  & 0.247 (0.040)  \\
  & &  biLSTM & 0.260 (0.054)  & 0.247 (0.048)  & 0.260 (0.043)  & 0.265 (0.046)  & 0.253 (0.085)  & 0.237 (0.089)  \\ 
& &   biLSTM+C & \textbf{0.221} (0.046)  & \textbf{0.209} (0.041)  & \textbf{0.221} (0.037)  & \textbf{0.223} (0.039)  & \textbf{0.214} (0.073)  & \textbf{0.201} (0.076)  \\
\cline{2-9}
&
BSS  &  RF & 0.390 (0.133)  & 0.365 (0.077)  & 0.397 (0.071)  & 0.342 (0.139)  & 0.330 (0.164)  & 0.324 (0.162)  \\ 
& &   RF+C & 0.463 (0.158)  & 0.470 (0.140)  & 0.482 (0.089)  & 0.424 (0.172)  & 0.440 (0.194)  & 0.419 (0.167)  \\
  & &  MLP & 0.289 (0.205)  & 0.348 (0.146)  & 0.215 (0.170)  & 0.260 (0.191)  & 0.264 (0.208)  & 0.200 (0.153)  \\ 
& &   MLP+C & 0.388 (0.187)  & 0.450 (0.127)  & 0.331 (0.150)  & 0.366 (0.186)  & 0.357 (0.204)  & 0.304 (0.165)  \\
 & &   SVM & 0.447 (0.145)  & 0.420 (0.134)  & 0.403 (0.129)  & 0.378 (0.109)  & 0.404 (0.117)  & 0.391 (0.097)  \\ 
& &   SVM+C & 0.529 (0.126)  & 0.493 (0.117)  & 0.492 (0.109)  & 0.461 (0.091)  & 0.420 (0.101)  & 0.507 (0.083)  \\
 & &   LSTM & 0.450 (0.139)  & 0.467 (0.128)  & 0.436 (0.103)  & 0.422 (0.145)  & 0.442 (0.139)  & 0.421 (0.101)  \\ 
& &   LSTM+C & 0.531 (0.118)  & 0.536 (0.110)  & 0.511 (0.088)  & 0.500 (0.146)  & 0.461 (0.121)  & 0.506 (0.086)  \\
  & &  biLSTM & 0.484 (0.112)  & 0.505 (0.105)  & 0.489 (0.089)  & 0.466 (0.097)  & 0.428 (0.154)  & 0.432 (0.166)  \\ 
& &   biLSTM+C & \textbf{0.592} (0.095)  & \textbf{0.581} (0.090)  & \textbf{0.566} (0.075)  & \textbf{0.550} (0.082)  & \textbf{0.504} (0.165)  & \textbf{0.613} (0.074)  \\
\enddata
\end{deluxetable*}

\vspace{-2cm}
\bibliographystyle{aasjournal}

\end{document}